\documentclass[fleqn,usenatbib]{mnras}

\usepackage{newtxtext,newtxmath}

\usepackage[T1]{fontenc}

\DeclareRobustCommand{\VAN}[3]{#2}
\let\VANthebibliography\thebibliography
\def\thebibliography{\DeclareRobustCommand{\VAN}[3]{##3}\VANthebibliography}

\usepackage{graphicx}	
\usepackage{amsmath}	

\title[Point and extended source detection and characterization]{DECORAS: detection and characterization of radio-astronomical sources using deep learning}

\author[S. Rezaei et al.]{
S. Rezaei,$^{1,2}$\thanks{E-mail: rezaei@astro.rug.nl}
J. P. McKean,$^{1,3}$
M. Biehl$^{2,4}$ and 
A. Javadpour$^{5}$
\\
$^{1}$Kapteyn Astronomical Institute, University of Groningen, Postbus 800, NL-9700 AV Groningen, the Netherlands\\
$^{2}$Bernoulli Institute for Mathematics, Computer Science and Artificial Intelligence, University of Groningen, Postbus 407, NL-9700 AK Groningen, the Netherlands\\
$^{3}$ASTRON, Netherlands Institute for Radio Astronomy, Oude Hoogeveensedijk 4, 7991 PD Dwingeloo, the Netherlands\\
$^{4}$SMQB, Institute of Metabolism and Systems Research IBR Tower, Level 3 College of Medical and Dental Sciences, University of Birmingham, Birmingham, UK\\
$^{5}$School of Mathematics, Computer Science and Technology, Guangzhou University, Guangzhou, China, 510006
}

\date{Accepted XXX. Received YYY; in original form 2021 August 30}

\pubyear{2021}

\begin{document}
\label{firstpage}
\pagerange{\pageref{firstpage}--\pageref{lastpage}}
\maketitle

\begin{abstract}
We present DECORAS, a deep learning based approach to detect both point and extended sources from Very Long Baseline Interferometry (VLBI) observations. Our approach is based on an encoder-decoder neural network architecture that uses a low number of convolutional layers to provide a scalable solution for source detection. In addition, DECORAS performs source characterization in terms of the position, effective radius and peak brightness of the detected sources. We have trained and tested the network with images that are based on realistic Very Long Baseline Array (VLBA) observations at 20 cm. Also, these images have not gone through any prior de-convolution step and are directly related to the visibility data via a Fourier transform. We find that the source catalog generated by DECORAS has a better overall completeness and purity, when compared to a traditional source detection algorithm. DECORAS is complete at the $7.5\sigma$ level, and has an almost factor of two improvement in purity at $5.5\sigma$. We find that DECORAS can recover the position of the detected sources to within $0.61\pm0.69$~mas, and the effective radius and peak surface brightness are recovered to within 20 per cent for 98 and 94 per cent of the sources, respectively. Overall, we find that DECORAS provides a reliable source detection and characterization solution for future wide-field VLBI surveys.
\end{abstract}




\section{Introduction}

Machine learning, and in particular deep learning, has been widely used for solving a number of astronomical problems (see \citealt{baron2019machine} for a recent review). This is because traditional approaches, such as visual inspection or model fitting, can be less effective when characterizing datasets that are growing in both size and complexity. Several machine learning frameworks can be applied in
this context, but two have gained the most attention in recent years: supervised and unsupervised learning.

In a supervised learning algorithm, the data acts as an instructor, assisting the model in discovering a relationship between a collection of features and user specified labels. Models
are trained to predict the properties of  unseen data \citep{Goodfellow-et-al-2016}. Supervised learning has been used to address various problems in astronomy, for example, the classification of galaxy morphologies in imaging data \citep{Will_mergers, Nolte_2019}, variable star classification using light-curve representation \citep{variable_star_classification}, and the classification of blazar candidates \citep{blazar_classification}. 

In the case of unsupervised learning, the algorithm simply receives data without target
labels or any feedback from the environment. The goal of the analysis is to identify patterns and structures within the data that are then used to make decisions, predict future inputs, or communicate inputs efficiently to another machine \citep{Ghahramani2004}. An example of such an algorithm is an autoencoder, which is a form of unsupervised convolutional neural network (CNN). Autoencoders have been used for a variety of astronomy applications, including real-time transient detection \citep{nima_transient} and the analysis of gravitationally lensed objects \citep{2017Hezaveh}. They have also been used for outlier detection; see for example \citet{gan_outlier}, in which a network is trained with normal data before computing an anomaly score for test query samples. As another example, \citet{outlier_supernova} presented a method for detecting rare transients or completely new flaring events of unknown physical nature. These are just a few of the many applications of supervised and unsupervised learning in astronomy. For a very brief review and further references, see for example,  \cite{esann-session}.

Here, we focus on a deep learning based approach to study astronomical images that have been made using radio interferometric techniques. In contrast to optical instruments, which capture images of the sky brightness distribution directly, radio telescopes employ interferometry to calculate the two-dimensional discrete intensity distribution of the sky, known as visibility data. A Fourier transform of the visibility data is then performed to produce an image of the sky. The result of this process is the convolution of the true sky brightness with the point spread function (PSF) of the interferometric array, which is commonly referred to as the dirty image. Due to the incomplete sampling of the interferometric visibility data, the PSF (also referred to as the dirty beam) has strong sidelobes that affect the entire image. This can make it difficult to recover the true sky brightness distribution from interferometric data. A common solution to this problem was presented by \citet{CLEAN}, who developed the {\sc clean} algorithm, which iteratively performs a deconvolution of the image by representing the underlying source brightness using simple parametric models, such as delta- or truncated Gaussian functions. The final step convolves the model of the source with the clean beam (a Gaussian function), which is then added to the Fourier transform of the residual visibilities. 

To characterize the object properties from interferometric images, several commonly used object detection algorithms have been developed (note that these are applied to images that have gone through a prior deconvolution process). {\sc PyBDSF} \citep{pybdsf}, {\sc blobcat} \citep{hales2012blobcat} and {\sc aegean} \citep{aegean}, are all examples of Gaussian fitting source detectors. {\sc ProFound} \citep{profound} on the other hand does not force any predefined parametric model to the detected sources, but is based on the segmentation of pixels in the neighborhood of the brightest pixel. The benefit of {\sc ProFound} over other source detection algorithms is the more accurate flux recovery for extended sources, as the detection is not based on a specific morphology. From an analysis of simulated observations that match the instrument properties of the Very Large Array (VLA), the completeness and reliability of {\sc PyBDSF}, {\sc aegean} and {\sc ProFound} for compact objects detected with a signal-to-noise ratio $> 4.3$ was found to be less than 85 per cent \citep{profound}.

Machine learning, and in particular, CNNs have already been widely used in the analysis of radio interferometric data. For example, they have been used to classify radio galaxies \citep{bowles2021attention}, to determine galaxy morphologies \citep{morphology_classification}, and to select pulsar candidates \citep{pulsars_CNN}. More specifically, CNNs have been employed to detect astronomical sources within the {\sc ConvoSource} \citep{ConvoSource}, {\sc DeepSource} \citep{deepsource} and Point Proposal Network (PPN; \citealt{tilley2020point}). Compared to traditional source detection algorithms, which can fail to detect sources when the signal-to-noise ratio is low, or can make false detections in regions of the images where the noise is highly correlated, the learning process in CNN-based source detectors has been shown to generate more accurate results. \citet{deepsource} and \citet{ConvoSource} have shown that CNN based source detection algorithms are more complete down to a signal-to-noise ratio of 4 in detecting compact sources when compared to {\sc PyBDSF}. However, these CNN based algorithms are all optimised for the analysis of images that have been deconvolved.
 
In this paper, we investigate a deep learning based object detection algorithm that characterizes the source properties from images that have not undergone any prior deconvolution. This is partly due to the complexity of applying deconvolution methods, like {\sc clean}, to large datasets. Furthermore, such an algorithm could in principle also be applied directly to the visibility data via a Fourier transform, which would remove the need for any imaging step. We focus our analysis on dirty images that are produced from a sparse radio interferometric array, namely the Very Long Baseline Array (VLBA). This is because we are interested in developing a new object detection algorithm that can be applied to wide-field Very Long Baseline Interferometric (VLBI) observations with instruments like the VLBA in the future. 
 
The presented approach for detection and characterization of radio-astronomical sources (DECORAS) using deep learning consists of four main steps. The first step uses encoder-decoder networks to remove the noise and dirty beam from the given dirty images (the Fourier transform of the observed interferometric visibility data). The predicted model images at the output of the encoder-decoder are used in the post processing step to find the position of the source. In the third step, another encoder-decoder is used to characterize the source structure. Finally, the extracted latent variables of the trained encoder-decoder network are used to recover the source surface brightness distribution. Unlike {\sc DeepSource} and {\sc PPN}, which have thus far only addressed the detection of unresolved objects, DECORAS is trained on both point and extended source detection and characterization. 

Our paper is arranged as follows. In Section~\ref{method_section}, the training/verification data and the detailed methodology of the source detection and characterization algorithm is presented. In Section~\ref{result-detection}, we evaluate our results by applying the algorithm to test data and compare with the results from using a traditional source detection code ({\sc blobcat}). In Section~\ref{result-characterization}, we investigate how well our algorithm can recover the source properties, such as source position, major axis and the true source surface brightness distribution. Finally, the results from this work are discussed and we present our concluding remarks on the methodology and future prospects in Section~\ref{discussion}.

\section{Method} \label{method_section}

This section presents our source detection and characterization methodology. First, the process of generating realistic simulated images is explained. This simulated dataset is used as the training and test samples for our network. Next, an overview of our approach is provided, with an explanation of the choice of loss function and specific network architecture that we have used. Then, we present our post processing object detection step, which determines the position of the source. Finally, our source characterization methodology is presented, which provides the structure and surface brightness distribution of the detected objects. 

\subsection{Simulating a representative training and testing dataset} \label{data}

Generating realistic images to train the network is one of the main steps for developing a source detection and characterization platform. This is because the network must learn the key properties and features of the data. Also, a simulated dataset can be used to test the robustness and completeness of the methodology, providing these data are unseen by the network during the training stage. Our goal is to develop a network that is applicable to data from sparse interferometric arrays, which are typical of VLBI observations. For our simulations, we have chosen to use the VLBA at an observing wavelength of 20 cm as our training and test dataset, the reasons for which we describe below. However, we see no obvious reason why our methodology cannot be used with other VLBI arrays that observe at other wavelengths, for example, the International LOFAR Telescope (ILT), the European VLBI Network (EVN), the Atacama Large Millimetre Array (ALMA), or the Square Kilometre Array (SKA-VLBI), which (will) operate from m to sub-mm wavelengths.

\begin{figure}
 \includegraphics[width=\columnwidth]{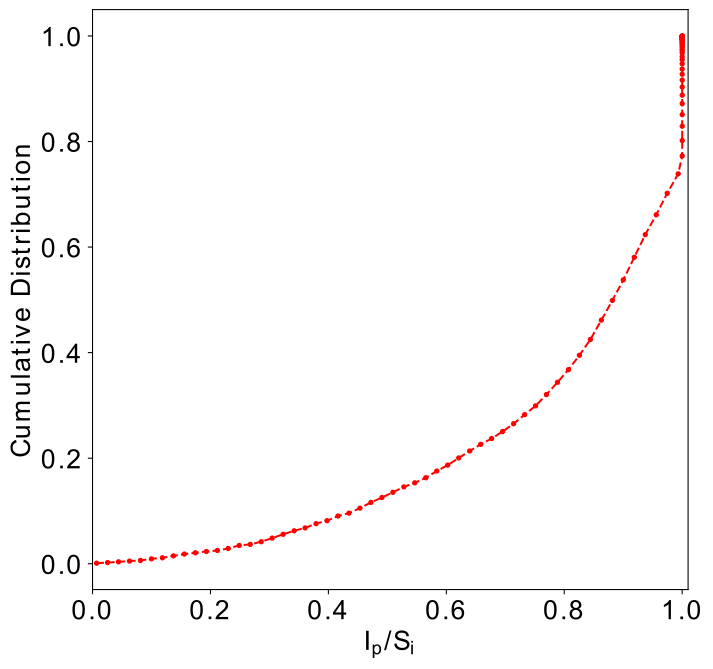}
 \caption{Cumulative distribution of peak surface brightness divided by the integrated flux density (in units of beam$^{-1}$) for detected objects in the mJIVE--20 survey \citep{Adam_mjive}. Around 20 percent of the detected sources in the mJIVE--20 survey have an equivalent peak surface brightness and flux density (unresolved sources). In more than 65 percent of all detected sources in the mJIVE--20 survey, the peak surface brightness is about 80 percent of the total flux density of the source, meaning that the majority of the sources detected in the mJIVE--20 survey are compact.}
 \label{fig:cumulativespsi}
\end{figure}

\begin{figure}
\begin{center}

 \includegraphics[width=\columnwidth]{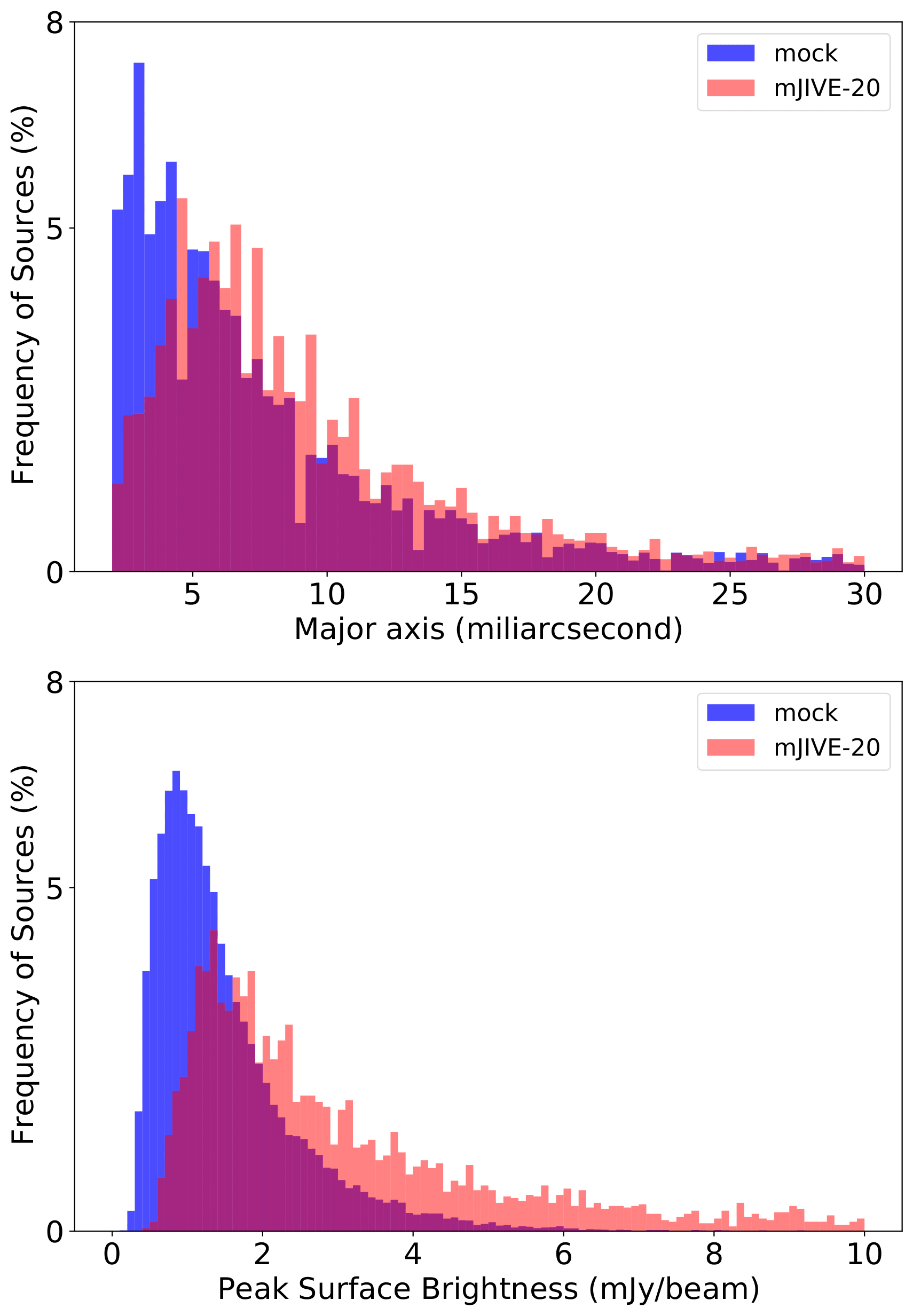}
 \caption{A comparison of the measured size and peak surface brightness of the injected mock and the real sources detected as part of the mJIVE--20 survey. The upper plot compares the distribution of fitted Gaussian major axis. The lower plot shows the distribution of peak surface brightness for the two populations. The implemented Monte Carlo approach based on the mJIVE--20 survey catalog, has made it possible to create a training dataset that follows the physical characteristics of real sources.}
 \label{fig:mockedvsmjive}
 \end{center}
\end{figure}

\begin{figure*}
 \includegraphics[width=\textwidth, trim = 2cm 15.25cm 2cm 1.5cm]{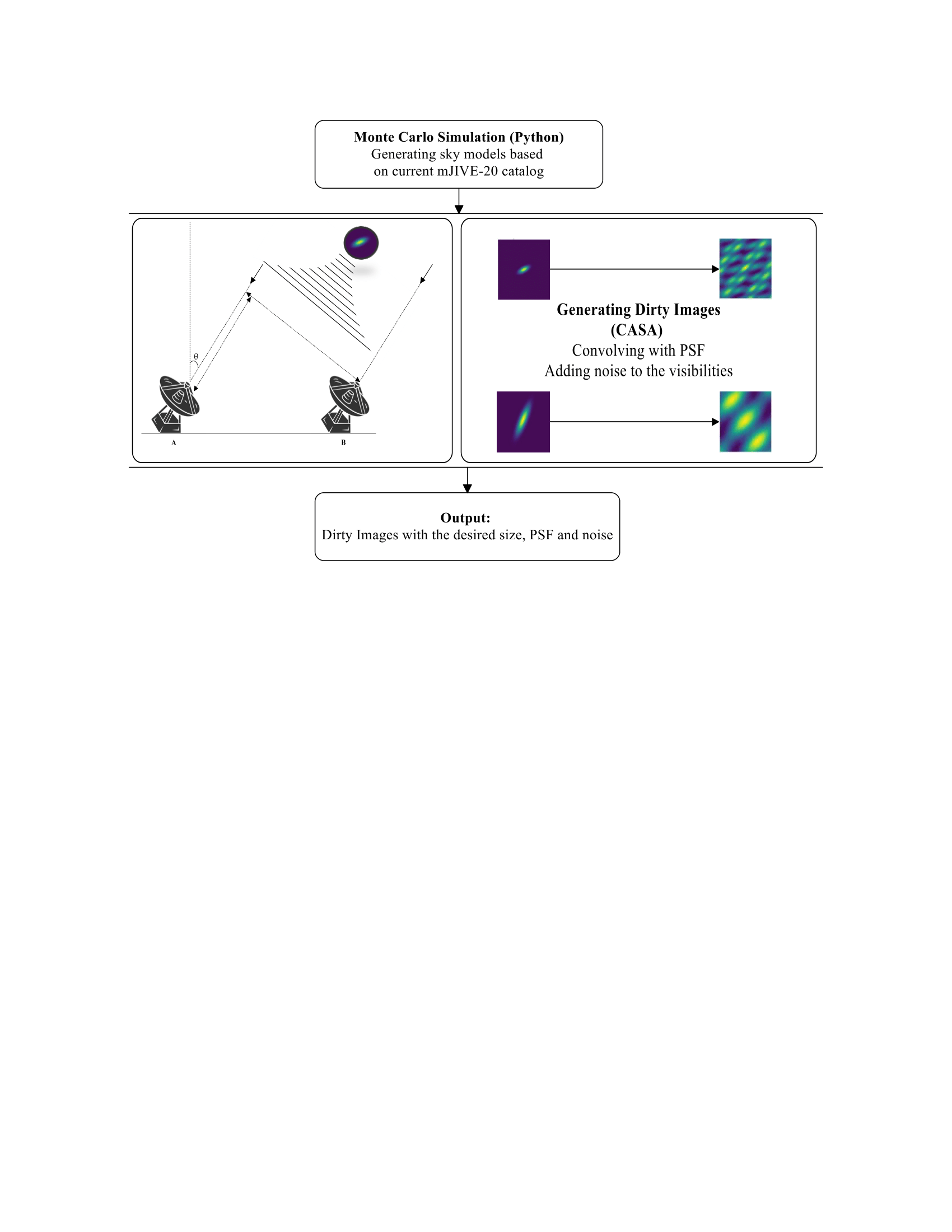}
 \caption{Flowchart of the mock data simulator, built within {\sc casa}. First, mock sources are generated using a Monte Carlo simulation to mimic the physical characteristics of the real sources in the mJIVE--20 survey. The generated sky model images are imported to {\sc casa}, using the simulator tool to generate the corresponding $uv$-datasets and dirty images.}
 \label{fig:sim_src}
\end{figure*}

We have chosen the VLBA as our test interferometer as it is a homogeneous array of ten 25-m radio telescopes, with separations between 236 and 8611~km. This means the generated datasets will sparsely sample the visibility plane (45 measurements per time and frequency interval), will have uniform and predictable noise properties, and will be $\sim2$~MB in size per simulated observation. By using the VLBA at 20 cm, we can also utilize actual observations from a large statistical sample of radio sources to provide realistic observational conditions and representative source models for our simulations. For this, we have used data from the mJy Imaging VLBA Exploration (mJIVE--20) survey \citep{Adam_mjive}, which targeted 24\,903 radio sources in an area of 238 square degrees from 306 unique observations. The total observing time was 1 h for each observation, which was further divided into a set of four sub-pointings around a bright calibrator source. Within $\sim20$~arcmin of the central calibrator, the rms noise in the deconvolved images is about 150 $\mu$Jy~beam$^{-1}$. The detection threshold for the mJIVE--20 survey was set to $6.75\sigma$ (based on simulations to determine the completeness), where $\sigma$ is the rms map noise of a given observation. This led to 4\,965 radio sources being detected in the deconvolved images using {\sc blobcat}. 

We have developed a framework with three steps to create mock training and test data with similar properties to those obtained by the mJIVE--20 survey. The first step defines the images in terms of angular size and number of pixels. The size of the individual pixels is set such that the visibility data are at least Nyquist sampled in the image plane, which for VLBA observations at 20 cm with a PSF sampling of four gives a pixel size of 1.25 mas. The majority of the detected sources in the mJIVE--20 survey are compact \citep{Adam_mjive}; here we define compactness ($C$) as the ratio of the integrated flux density ($S_i$) and peak surface brightness ($I_p$), where those objects with $C<1.25$~beams are classified as being compact. Fig.~\ref{fig:cumulativespsi} shows the cumulative distribution of the peak surface brightness to integrated flux density ratio of detected sources in the mJIVE--20 survey, which demonstrates that the majority of the sources are unresolved. Considering the characteristics of the detected sources in the mJIVE--20 survey, we have chosen the size of the input image to be $256\times256$~pixels, which is equivalent to an angular size of $320\times320$~mas on the sky. Larger images would increase the number of learning parameters in our encoder-decoder model. This means that the learning procedure would require more time and memory. 

Our simulated datasets are made using the Common Astronomy Software Applications ({\sc casa}; \citealt{CASA}) package and custom-written Python scripts. The mock sources were generated using either delta or Gaussian functions, with a defined peak surface brightness, size, ellipticity and position angle, and a random position within each image. The defined source properties were determined from a Monte Carlo simulation of all of the sources observed as part of the mJIVE--20 survey. Fig.~\ref{fig:mockedvsmjive} shows a comparison of the injected mock sources and the actual mJIVE--20 survey sources in terms of their size (major axis) and peak surface brightness. Note that the number of injected mock sources is greater than the actual number of detected sources in the mJIVE--20 survey, as we have injected mock sources to all of the 24\,903 phase-centres, while the catalog only contains information for 4\,965 sources. We have also included fainter sources in the generated mock data to test our detection algorithm for sources with a low signal-to-noise ratio. The output of this step is the sky model interferometric datasets which will be used to generate dirty images.

{\sc casa} stores interferometric visibility data in a format called MeasurementSet. While it is possible to create a simulated MeasurementSet from  scratch, the {\sc casa} simulator tool can use an existing MeasurementSet to obtain the position of the antennas and other observational settings (frequency and time sampling). Using actual MeasurementSets is a good choice for this work as we aim to train our network with mock data that is representative of real data. In this way, the simulator samples data with the correct $(u,v)$ coordinates, considering the current model image with the mock source in it. We have used the existing MeasurementSets of the actual mJIVE--20 survey observations to generate simulated visibility datasets. To take the thermal noise into consideration, we have added Gaussian noise to the visibility data that is representative of the noise properties of the mJIVE--20 survey. We have not included any systematic errors to the real or imaginary component of the visibilities. Finally, the dirty image is generated as the result of taking the Fourier transform of the visibility data, and gridding using the pixel size and number of pixels described above. Fig.~\ref{fig:sim_src} presents a flowchart that summarizes how the simulated dataset of model and dirty images is generated using  {\sc casa}. 

\begin{figure}
\begin{center}

\begin{minipage}[b]{\columnwidth}
\centering
\includegraphics[height=3cm]{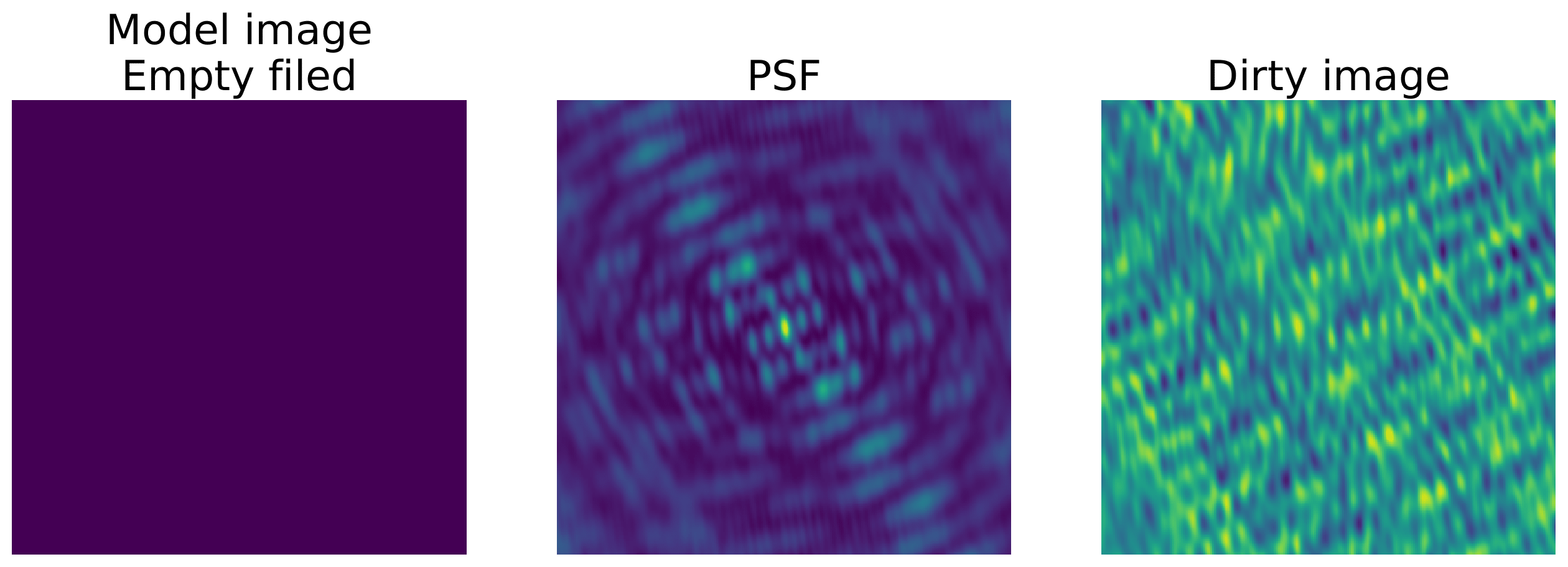}
\end{minipage}

\begin{minipage}[b]{\columnwidth}
\centering
\includegraphics[height=3cm]{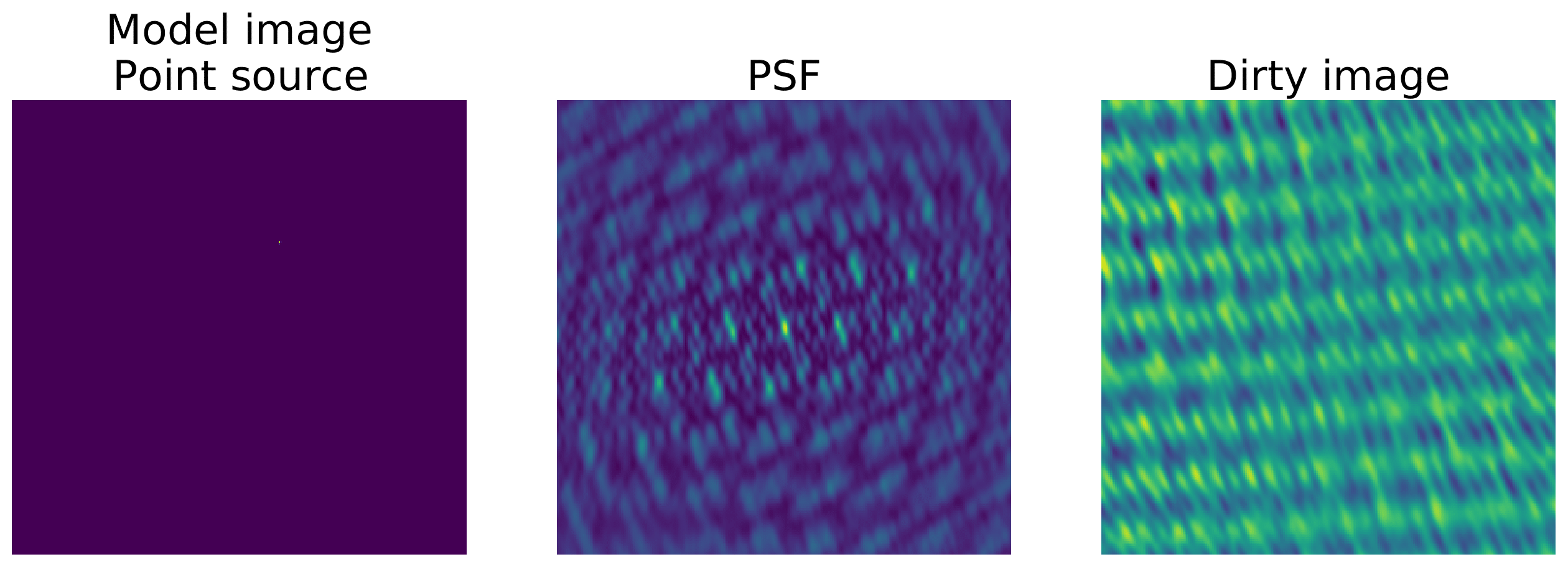}
\end{minipage}

\begin{minipage}[b]{\columnwidth}
\centering
\includegraphics[height=3cm]{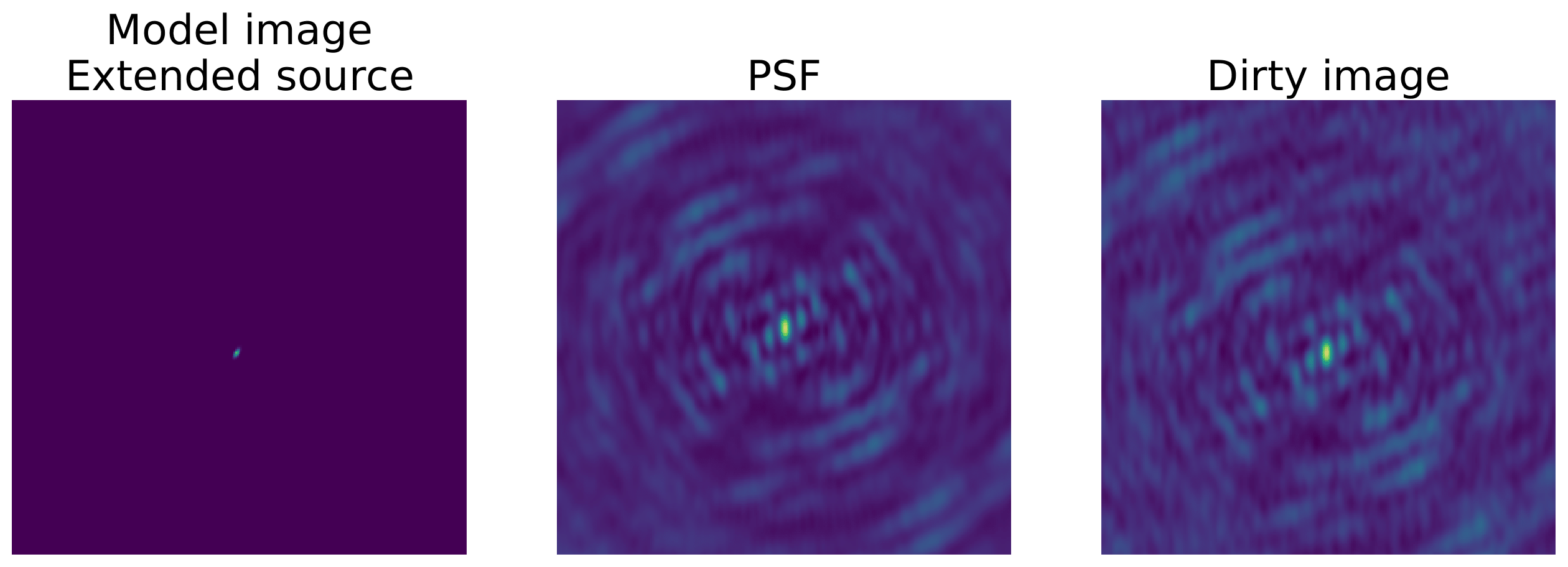}
\end{minipage} 

\begin{minipage}[b]{\columnwidth}
\centering
\includegraphics[height=3cm]{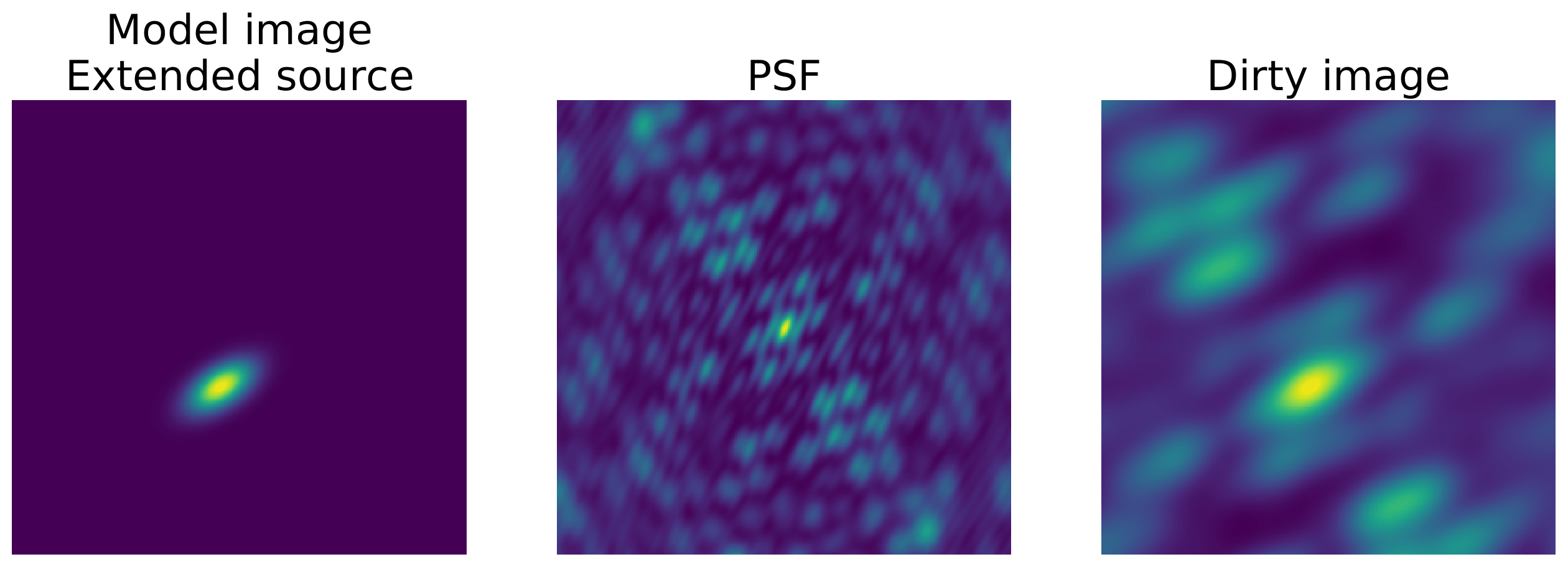}
\end{minipage} 
\newpage
\caption{Some examples of the generated model images (left), PSFs (middle) and the final dirty images (right) fed to the network. The model images are the output of the implemented source generator. The PSFs are determined from the Fourier transform of the $uv$-sampling function, taken from the original mJIVE--20 survey observations. The dirty images are the final output of the source simulator. Each image contains $256\times256$~pixels and is equivalent to a sky-area of $320\times320$~mas$^2$.}
\label{fig:allselectedsources}

\end{center}
\end{figure}

In total 50\,000 simulated images, from 306 observations were made for training the network. In Fig.~\ref{fig:allselectedsources}, we provide a few examples of the model sources, the PSF of the individual observations, and the final dirty images used to build the training and validation dataset. The dirty images are the input to the network, while the model images are the output of our encoder-decoder network.

\subsection{Overview of DECORAS}
\begin{figure*}
 \includegraphics[width=\textwidth, trim= 2.5cm 16cm 4.75cm 2.5cm]{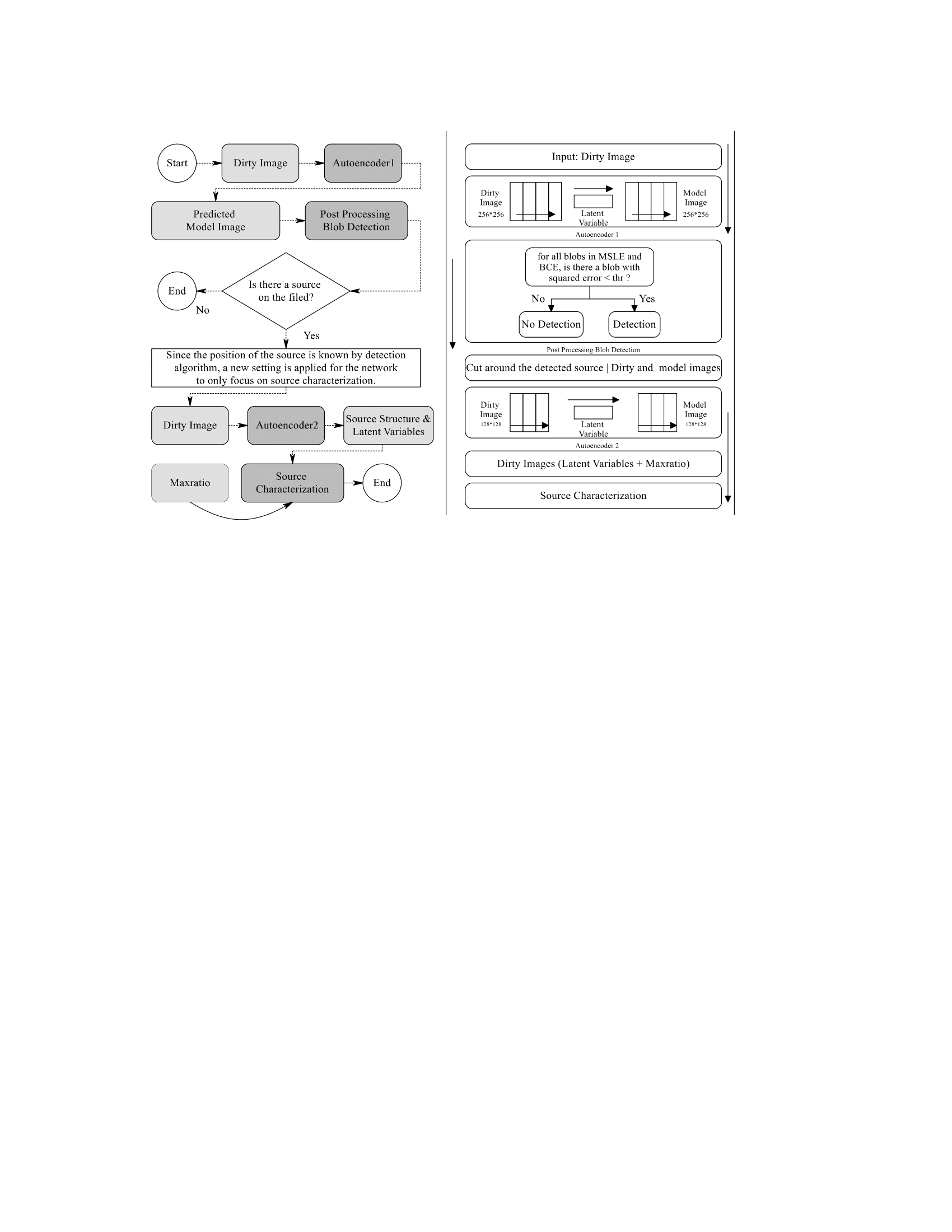}
 \caption{The left panel shows the general flowchart of DECORAS. The grey rectangles in the flowchart represent the functions or the developed algorithms in which the source detection and characterization are designed. The transparent rectangles show input/output data. In the right panel, further details on each of the grey boxes in the flowchart are given. A more detailed view of Autoencoder1 is represented in Fig.~\ref{fig:encoder_decoder_structure}, where four convolutional layers that transforms dirty images of $256\times256$ pixels into the model images of the same size are shown. The Post Processing Blob Detection is responsible for locating the injected source in the predicted model image. To characterize the physical properties of the source, Autoencoder2 has been used.  It has the same network structure as Autoencoder1, except the input shape is $128\times128$ pixels, and therefore, the number of convolutional layers is smaller. The source characterization uses the generated latent variable and predicted model images by Autoencoder2 as well as the maxratio to estimate the source surface brightness distribution.}
 \label{fig:flowchart_diagram}
\end{figure*}

 Given any input dirty image, DECORAS is trained to deconvolve the PSF, remove thermal noise, locate the possible source in the image and characterize the source structure and surface brightness distribution. This process is summarised in the flowchart presented in Fig.~\ref{fig:flowchart_diagram}. 
 
 The first step of DECORAS consists of encoder and decoder parts. It is trained to recognize beam effects, correlated noise, or other sources of contamination in the dirty image, and to remove them from the predicted output. The first step of DECORAS is hereafter referred to as Autoencoder1. The output of Autoencoder1 is then passed to the Post Processing Blob Detection function to find the position of the source in the predicted model image. At the end of this step, the existence of the source and its position are known to the algorithm. To investigate the physical characteristics of the source, we crop around the region of interest in the field to form a smaller model and dirty image, with the source at the centre. The cropped images are fed to  Autoencoder2 to investigate the physical characteristics of the detected source. Our experiments show that using Autoencoder2 yields a higher accuracy than with Autoencoder1 when extracting physical properties of the underlying source. Autoencoder2 has the same basic structure as Autoencoder1. However, due to the smaller image size, Autoencoder2 requires fewer convolutional layers in the encoder and decoder parts. 
 
 The right panel of Fig.~\ref{fig:flowchart_diagram} provides more information on the grey boxes shown in the left panel. For example, Autoencoder1 and Autoencoder2 are shown with four and three symbolic convolutional layers for the encoder and decoder, respectively. For the post processing step, the detection strategy is provided. More information about the Post Processing Blob Detection is presented in Section~\ref{postprocessingblobdetection}.

\subsection{Preprocessing}
It is important to keep the pixel values of all the images in the same range of (0,1) in order to optimize the learning process while minimizing the achieved loss in the network. The process of normalization used here linearly transforms the pixel values on all the images to a common range of between (0,1). We have used a MinMax normalization according to,
\begin{equation}
\centering
    x_{\rm normalized}= \frac{x-\min(x_d)}{\max(x_d)-\min(x_d)},
\label{eq:minmax}
\end{equation}
 where $x$ is the value of a given pixel and $x_d$ represents all the pixels in the image.

As we will show below, DECORAS performs very well when the pixel values in each image are normalized in this way. However, the normalized predicted model image is not useful for recovering the absolute surface brightness of the detected sources. We have addressed this problem by analysing the latent variables generated by Autoencoder2 in the training process. This method is based on the information that the network has captured through the learning by removing the thermal noise and deconvolving the dirty images to obtain the expected model images. We discuss the details of estimating the surface brightness of the detected sources in Section \ref{SBE}.

\subsection{Network structure}

\begin{figure*}
 \includegraphics[width=\textwidth]{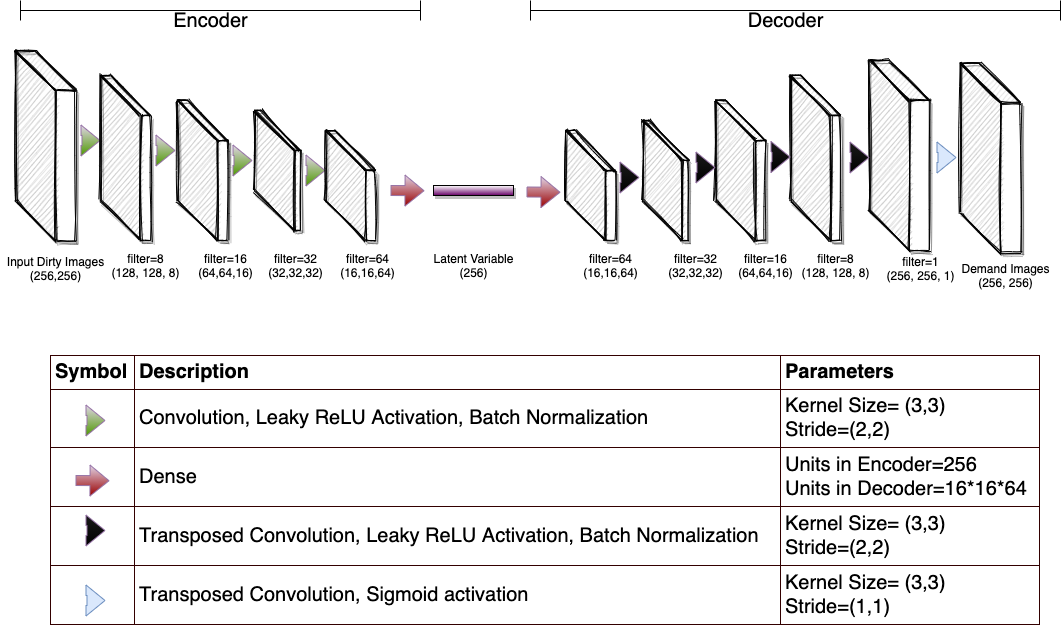}
 \caption{ The encoder-decoder network structure used in Autoencoder1 for source detection. There are four convolutional layers on both the encoder and decoder sections. Each convolutional layer is followed by leaky ReLU activation and batch normalization. In the encoder section, a stride of (2,2) is defined for the convolutional layers, which are used to down sample the data. At the end of the encoder, a dense layer is used to transform the extracted set of features in the convolutional layers into the latent variables. The decoder has a similar structure, but instead of down sampling the input, it up samples the data using transposed convolutional layers. A sigmoid activation function is placed at the end of the decoder to ensure the output pixel values are between (0,1).}
 \label{fig:encoder_decoder_structure}
\end{figure*}

Fig.~\ref{fig:encoder_decoder_structure} presents the architecture of the network, which consists of two main components: an encoder and decoder in which convolution, leaky ReLU (Rectified Linear Unit) activation and batch normalization are used sequentially. We use convolution with a stride of (2,2), which down samples the input image by a factor of 2 in each axis. A fully connected neural network is placed at the final step of our encoder. It learns the weights of the neurons for producing 256 values of latent variables. The latent variables that are generated by the encoder are the only piece of information our decoder uses to determine the output model image. This means that the compression rate of the encoder that imports the dirty images and generates the latent variables is 256. This is because the network only takes the structure and position of the sources into account. All of the other information about the  correlated noise, PSF and beam effects are learnt to be ignored by the network. In forthcoming work we will implement networks that can also solve for the PSF, and hence determine if there are residual calibration errors in the data. The same process, using the same parameter variables, is applied to the decoder part, but in a reverse order. Instead of using a convolution, the decoder uses the transpose convolution with a stride of (2,2) to up-sample the input image by a factor of 2. Leaky ReLU and batch normalization are also included, as shown by the grey arrows in Fig.~\ref{fig:encoder_decoder_structure}. After obtaining the desired image size by up-sampling, an extra layer of convolution and sigmoid activation is used to force the final generated pixel values to be between 0 and 1.

Our choice of network structure and the given input/output images can be considered as some sort of image segmentation. In such algorithms, the input image goes through different layers of convolution, and the generated latent variables are used to remap to a full output image. Instead of reconstructing the input image (the typical use of autoencoder structure), it only targets a specific segment of the image and, specifically for our case, where the source is located. Beside the location of the source, our network is sensitive to the size and structure of the source. On the other hand, as our network uses dirty images as the input and generates model images at the output, neurons on the encoder are forced to remove the effect of the PSF and correlated noise in the dirty images.

As described above, using a convolution layer with a stride of (2,2) down samples the input image by a factor of 2. Another alternative is to do so using a max pooling layer. We have chosen to use a stride parameter of (2,2) over the max pooling layer for two main reasons. First, the convolution layer has a learnable nature while the max pooling is a fixed function that takes the maximum value of each defined filter. Second, applying a stride parameter of (2,2) reduces the complexity of our algorithm as the convolution layer becomes cheaper to compute, and we do not need to add an extra layer of max pooling at the end of each major step in our encoder and decoder.

 A batch normalization is added at the end of each convolution layer to stabilize the distribution of inputs (over a minibatch). This is achieved by keeping the mean ($\mu_{x_k}$) and standard deviation ($\sigma$) of the output close to 0 and 1, respectively. It also helps to decrease the importance of the weight initialization and regularizes the model \citep{santurkar2018does}. In the Keras implementation, the output of a batch normalization layer is applied to each feature $x_k$, such that,
\begin{equation}
\centering
\hat{x}_k= \gamma\frac{x_k - \mu_{x_k}}{\sigma^2(x_k) + \epsilon} +\beta,
\label{eq:batchnorm}
\end{equation}
where $\epsilon$ is a constant that is being added to make sure the denominator is nonzero, $\gamma$ (initialized as 1) and $\beta$ (initialized as 0) are learnable parameters used for scaling and shifting purposes.

\subsection{Loss function}

In order to complete the learning process, it is necessary to define a loss function. The network uses the loss function to calculate the error between the estimated and expected model images at the end of each iteration. This error is used to update the weights and minimize the final error using optimizing strategies such as gradient decent. We have compared the performance of four different loss functions: Mean Squared Error (MSE), Mean Absolute Error (MAE), Binary Cross Entropy (BCE) and Mean Squared Logarithmic Error (MSLE). 

Our tests found that the MSE and MAE loss functions are not a proper choice for our specific problem since they have difficulties in detecting the source when the demand image is a point source (characterized by a single non-zero pixel while the rest of the image is zero). In order to force the network to take into account the information provided by a single pixel, \citet{deepsource} suggest to smooth the adjacent pixels around the point source to provide more non-zero pixels. On the other hand, \citet{nima_transient} have provided a solution that conditionally boosts the error on the non-zero pixels in the image. With this technique, the learning rate is virtually increased for only the non-zero pixels. Note that increasing the general learning rate is not a solution to the problem here as the learning procedure would generate a sub-optimal set of weights or an unstable training process. Based on the preliminary results of using the three defined loss functions with our network structure (see Fig.~\ref{fig:encoder_decoder_structure}), and the issues that MSE and MAE have when the source structure is very small compared to the size of the image, we have decided to only consider the MSLE and BCE further. 

Our encoder-decoder network is designed to solve a denoising problem. This can be interpreted as a pixel-wise classification in which each pixel (in the normalized images) is assigned a range between (0,1). We consider a given model image $M$ and a predicted model image $M'$ with size of $n \times n \times 1$. Considering $x_{i,j}$ as the pixel value on the $i,j$ position of the model image and $x'_{i,j}$ as the predicted pixel value at the same location, the BCE and MSLE are calculated according to
\begin{equation}
\centering
\begin{split}
{\rm BCE}\,(M',M)= -\frac{1}{n^2} \sum_{i=1}^{n}\sum_{j=1}^{n} x_{i,j}\log x'_{i,j} + \\
(1-x_{i,j})\log (1-x'_{i,j}),
\end{split}
\label{eq:bce}
\end{equation}
and
\begin{equation}
\centering
\begin{split}
{\rm MSLE}\,(M',M)= \frac{1}{n^2} \sum_{i=1}^{n}\sum_{j=1}^{n} \left[\log(x_{i,j})-\log(x'_{i,j})\right]^2,
\end{split}
\label{eq:msle}
\end{equation}
respectively. Note that the MSLE loss is similar to the MSE, but instead of calculating the loss of $M$ and $M'$ by computing the average, it calculates the squared error between the logarithm of the true and predicted values. Compared to MSE, MSLE penalizes underestimation more than overestimation.

\subsection{Post processing blob detection} \label{postprocessingblobdetection}
The network is designed to generate predicted model images with zero background, that is, the only non-zero pixel values in the predicted model images should be related to a detected source. In some cases, and particularly in the low signal-to-noise ratio regime, the network can get confused on where the input source is located and instead generates images with multiple components with non-zero pixel values (hereafter referred to as blobs). To identify such blobs in the images, we have used the {\sc Scikit} Image library \citep{van2014scikit} in Python, which contains several blob detection solutions. We have used the blob\_{dog} function to find blobs in the given predicted model images. The output of this algorithm is the $x$ and $y$ co-ordinates of any detected blobs, with an estimate of the uncertainty from the standard deviation of a fitted Gaussian function to the blobs. Through this process, we are able to catalog the candidate sources that have been identified by the network. 

\subsection{Surface brightness estimator} \label{SBE}

Losing the exact pixel values in the predicted model images due to the normalization process discussed above requires an alternative solution to estimate the absolute source surface brightness. To characterize the detected sources, DECORAS  relies on two sources of information. The first is the compressed set of features that are represented by the latent variables of our encoder-decoder structure. The latent variables are the only information that the decoder uses to construct the model images with all the embedded details, like source position and the physical characteristics. Fig.~\ref{fig:encoder_decoder_structure} shows the position of the latent variables in our encoder-decoder structure. In general, a greater number of latent variable units will result in a more clear and sharper reconstruction of the output image.

\begin{figure}
 \includegraphics[width=\columnwidth]{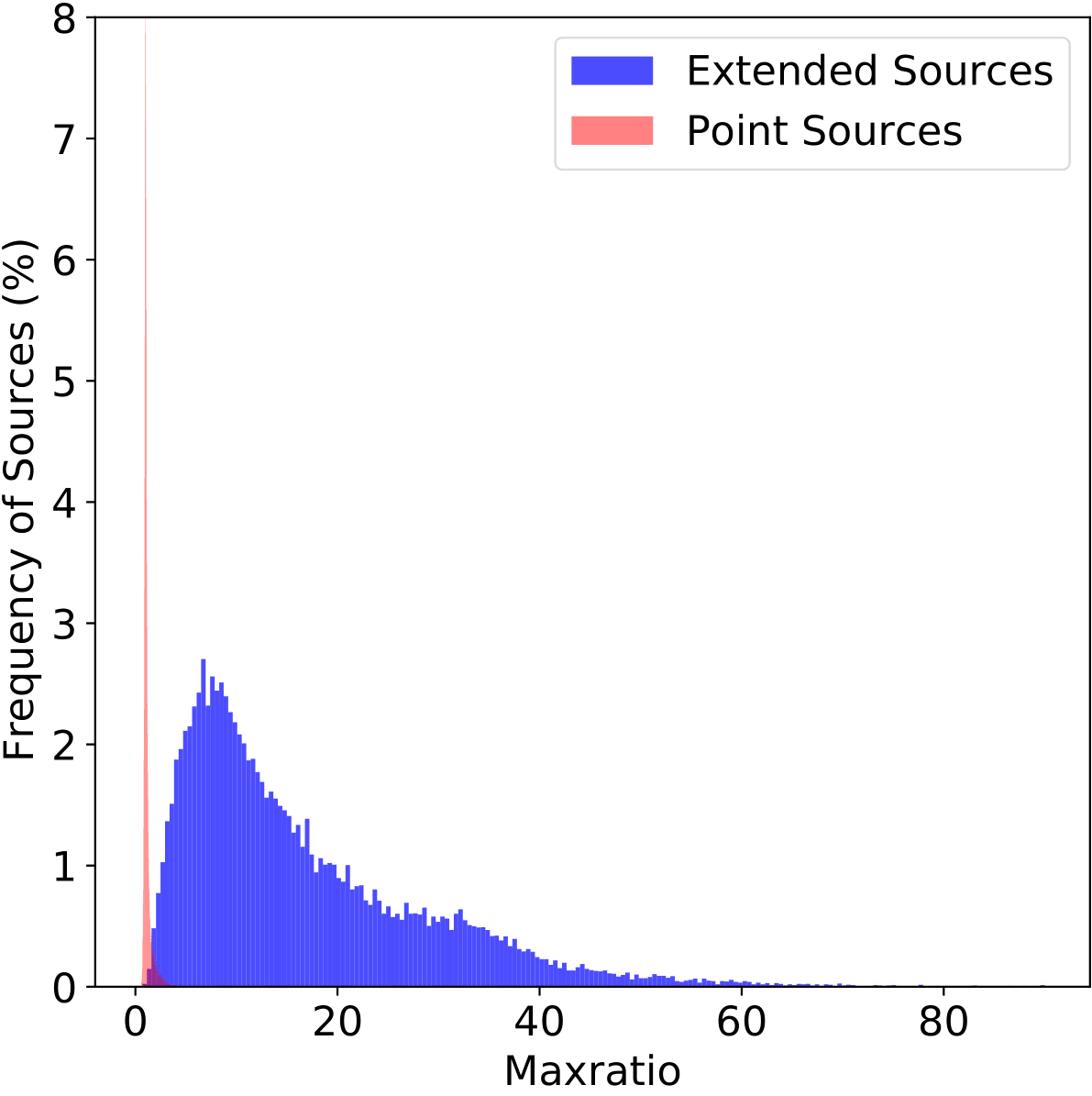}
 \caption{The maxratio distribution for point and extended sources. The maxratio is defined as the ratio between the peak surface brightness in any dirty image to its corresponding model image. It is used to estimate the absolute surface brightness of the detected sources.}
 \label{fig:maxratio}
\end{figure}

The second source of information is based on the fact that the source peak surface brightness in the dirty image is not equal to the peak in the model image. This happens due to the process of converting the visibility data of the true sky model to the dirty images (by adding Gaussian noise and convolving with the dirty beam). Convolving the visibility data of extended sources with the PSF increases the peak surface brightness in the predicted model image. The level of increase is dependent on the source size, PSF structure and the noise of the visibilities. We have measured this increase by calculating the ratio between the source peak surface brightness in the dirty image to the peak surface brightness of the corresponding injected source in the model image. This parameter, which we call the maxratio, is defined as
\begin{equation}
\centering
\begin{split}
{\rm maxratio} = {I_d}\big/{I_m},
\end{split}
\label{eq:maxratio}
\end{equation}
where $I_m$ is the peak surface brightness of the source in the model image (in units of Jy~pixel$^{-1}$) and $I_d$ is the peak surface brightness in the dirty image (in units of Jy~beam$^{-1}$). The maxratio is a measure of how much the peak surface brightness of a source has changed as it is convolved with the dirty beam to form the dirty image. Fig.~\ref{fig:maxratio} shows the distribution of maxratio for the entire simulated dataset. For point sources, the maxratio is mostly close to unity, as expected, and for extended sources the maxratio is typically larger, up to a factor of 90.

\begin{figure*}
 \includegraphics[width=\textwidth]{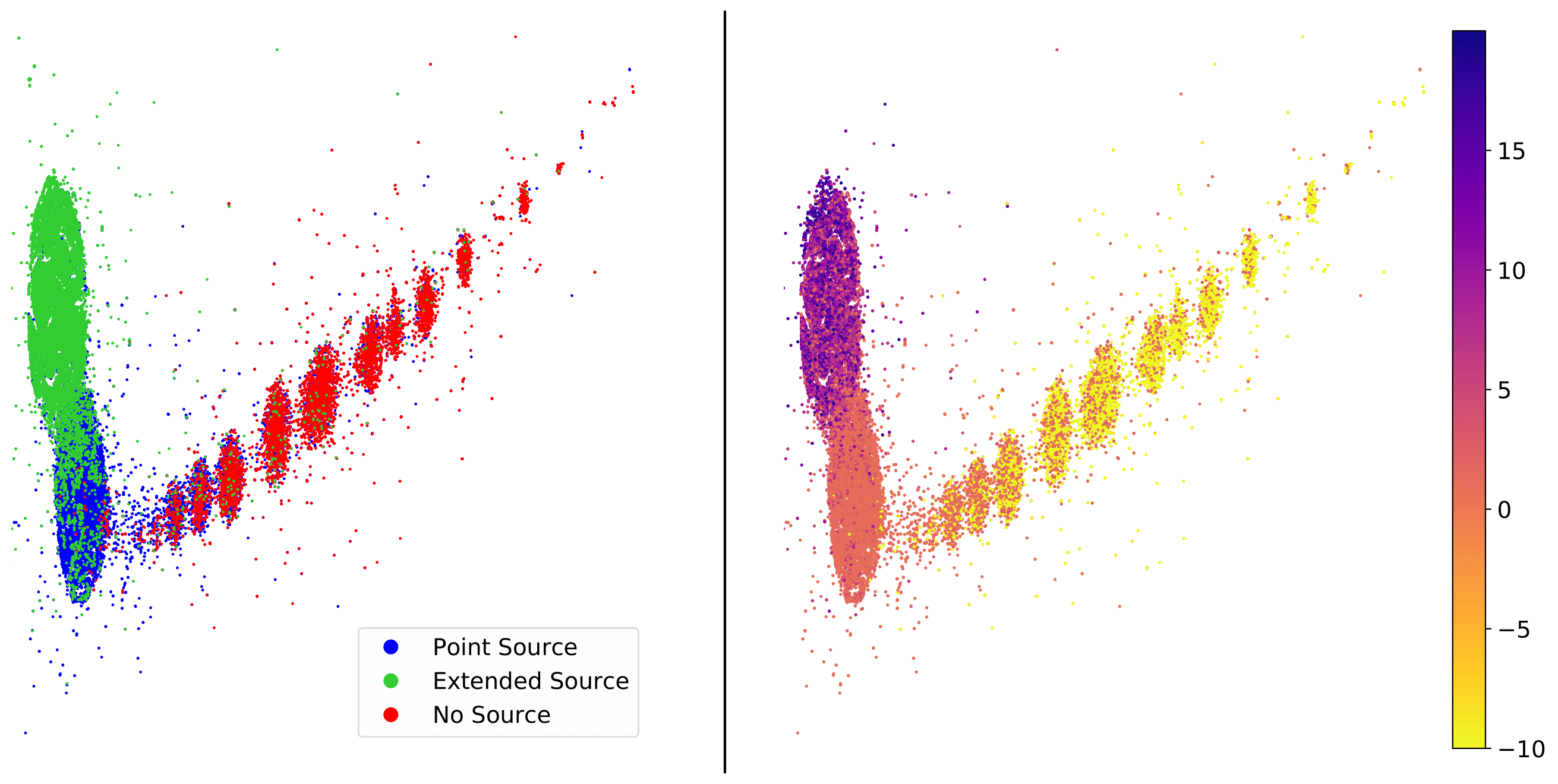}
 \caption{A t-SNE visualization  of the 256 latent variables  extracted from the encoder section of Autoencoder1. The encoder provides the compressed representation of any corresponding model image to the corresponding input dirty image. Latent variables carry the information of the source structure and accordingly the maxratio. The left panel is colour-coded by the class of data (point source, extended source, and noise realization). The right panel shows the same 2D representation of the latent variables coloured according to the corresponding maxratio. For the noise realization samples, with no injected source, the value of the maxratio has been assigned to $-10$ for visualization purposes.}
 \label{fig:TSNE_latent_MSLE}
\end{figure*}

A 2D visualization of the latent variables is a powerful tool to gain insight to the structure of data. One way to visualize high-dimensional latent variables is the t-Distributed Stochastic Neighbour Embedding (t-SNE; \citealt{van2008visualizing}) method. Fig.~\ref{fig:TSNE_latent_MSLE} presents the t-SNE visualization of our simulated data with two different colouring schemes. TSNE is implemented using {\sc scikit} Python package with the perplexity of 20 and the maximum number of iterations of 400. Each data point in Fig.~\ref{fig:TSNE_latent_MSLE} represents an array of 256 values, which are the extracted latent variables of Autoencoder1. The left plot of Fig.~\ref{fig:TSNE_latent_MSLE} is coloured according to the classes of data: point and extended source samples, and noise realization samples with no injected source. In the right plot of Fig.~\ref{fig:TSNE_latent_MSLE}, we have coloured the data points according to their corresponding maxratio. It is clear that the different source classes and their corresponding maxratios are separable using the latent variable information. Also, by determining the maxratio in this way, the absolute peak surface brightness of a given source can be recovered using equation~\ref{eq:maxratio}.

Although Fig.~\ref{fig:TSNE_latent_MSLE} provides some insight to the maxratio distribution and the types of sources that are detected, there is clearly some overlap between the three classes of data. Using several regression estimator techniques, such as the $k$-Nearest Neighbour (KNN), XGBoost and RandomForest, we find that this overlap affects the accuracy of the source brightness estimation. Therefore, we apply a specific approach for this task; once a source is detected and located by the network, its structure is inferred more accurately when the source is positioned in the centre of a cropped image. As the structure of the source is correlated with the maxratio (see Fig.~\ref{fig:maxratio}), this results in a more accurate prediction of the maxratio.

\begin{figure*}
 \includegraphics[width=\textwidth]{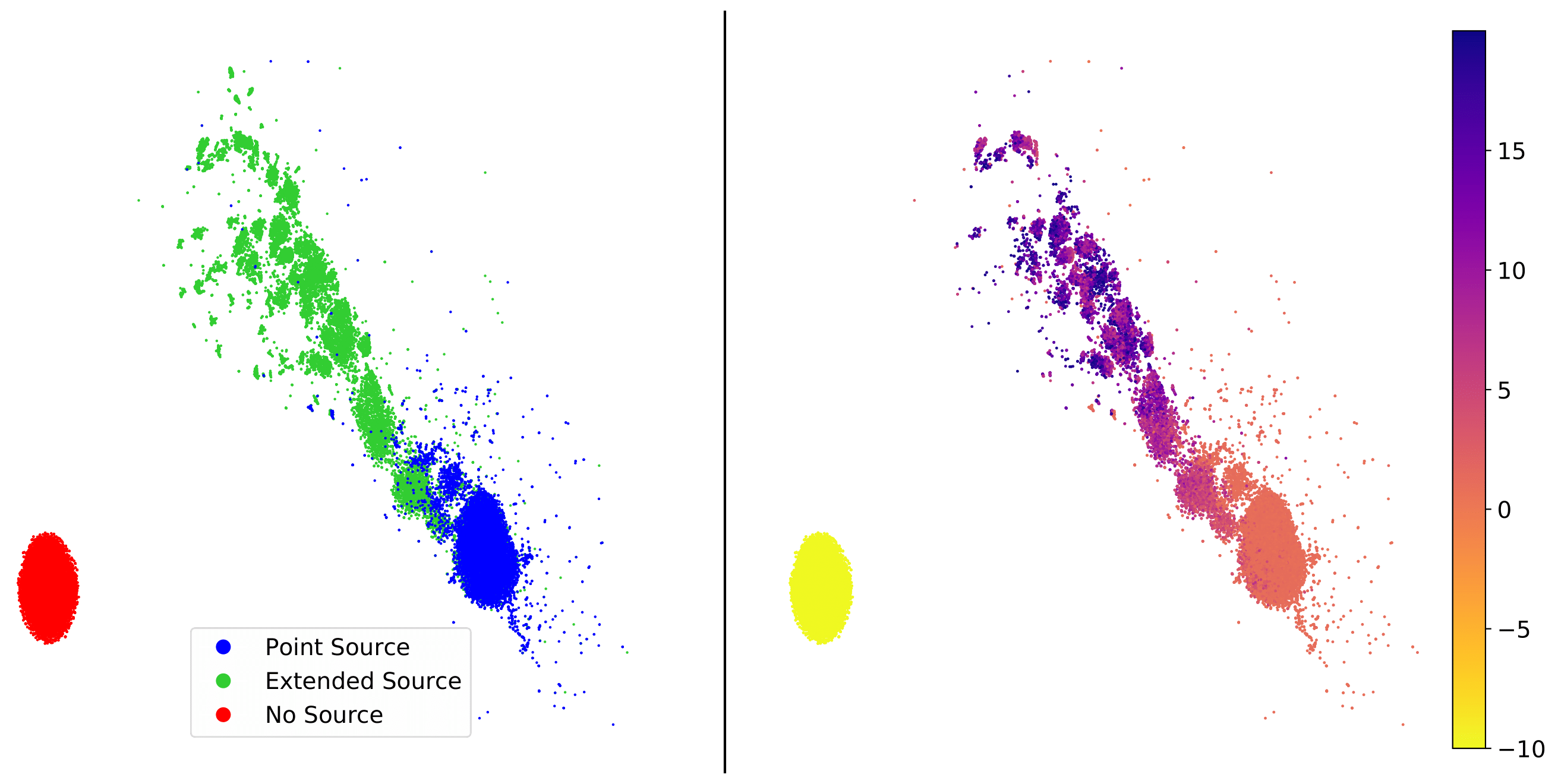}
 \caption{A t-SNE visualization of the latent variables extracted from Autoencoder2, which converts dirty images of $128\times128$ pixels into model images of the same size. The position of the injected source is the centre of the image. In the left panel, the colour illustrates the class of data (point source, extended source, and noise realization). The right panel shows the same 2D visualization of the latent variables coloured according to the corresponding maxratio. For the noise realization samples, with no injected source, the value of the maxratio has been assigned to $-10$ for visualization purposes.}
 \label{fig:TSNE_latent_MSLE_128}
\end{figure*}

In our implementation, a square image with $128\times128$~pixels is cropped around the position of the detected source in both the dirty and model images. These images are then used to train a new network (Autoencoder2), the latent variables of which provide a more accurate representation of the source structure. Fig.~\ref{fig:TSNE_latent_MSLE_128} shows the two-dimensional t-SNE visualization of the latent variables obtained from the training data where the source is centred and the image is cropped. A comparison of Figs.~\ref{fig:TSNE_latent_MSLE} and \ref{fig:TSNE_latent_MSLE_128} shows that the latent variables with the new setting yields 
 a more distinct representation of the different classes and the maxratio.

The network architecture used for the structure estimator is similar to the encoder-decoder shown in Fig.~\ref{fig:encoder_decoder_structure}. However, due to the smaller input image size, one convolutional layer from both the encoder and decoder sections is removed. In this reduced architecture, a layer of 64 units represents the latent variables.

\section{Source detection} \label{result-detection}

In this section, we present our results on training the network using the BCE and MSLE loss functions, before providing an overview of our DECORAS source detection strategy. The results from using DECORAS are compared to {\sc blobcat}, a traditional source detection algorithm, which was also used by the mJIVE--20 survey. 

\subsection{Defining the true positive and true negative rates}

In order to evaluate our results, we consider the confusion matrix presented in Table~\ref{tab:sample_confusion}. The true positives (TP) are defined as the number of fields with an injected source that the algorithm has successfully detected, while the true negatives (TN) are the number of fields with no injected source where the algorithm correctly returns a non-detection. Conversely, the false positives (FP) are defined as the number of fields with no injected source, but the algorithm detects a source, and the false negatives (FN) correspond to the number of fields with an injected source that the algorithm fails to detect. 

We quantify the performance of DECORAS and {\sc blobcat} when applied to simulated data by calculating the TP and TN rates, 
\begin{equation}
\centering
\begin{split}
{\rm  TP~rate}~= \frac{\rm TP}{\rm TP+FN},
\end{split}
\end{equation}
and
\begin{equation}
\centering
\begin{split}
{\rm TN~rate}~= \frac{\rm TN}{\rm TN+FP},
\end{split}
\end{equation}
respectively. Given that we aim to limit the number of FP and FN events, an ideal system would achieve a TP rate (sample completeness) and TN rate (sample purity) of 1.0.

The simulated test dataset consists of two components. First, we generate 8000 images in which 3000 samples correspond to point sources and 5000 samples contain (extended) elliptical Gaussian sources. To measure the signal-to-noise ratio for each source, we divide the peak surface brightness of the injected source in the model image by the rms of a noise realization of the same simulated observation without any injected source. For our simulations, we use a signal-to-noise ratio of the injected sources between 1 to 16. This simulated dataset is used to determine the TP rate.

\begin{figure}
 \includegraphics[width=\columnwidth]{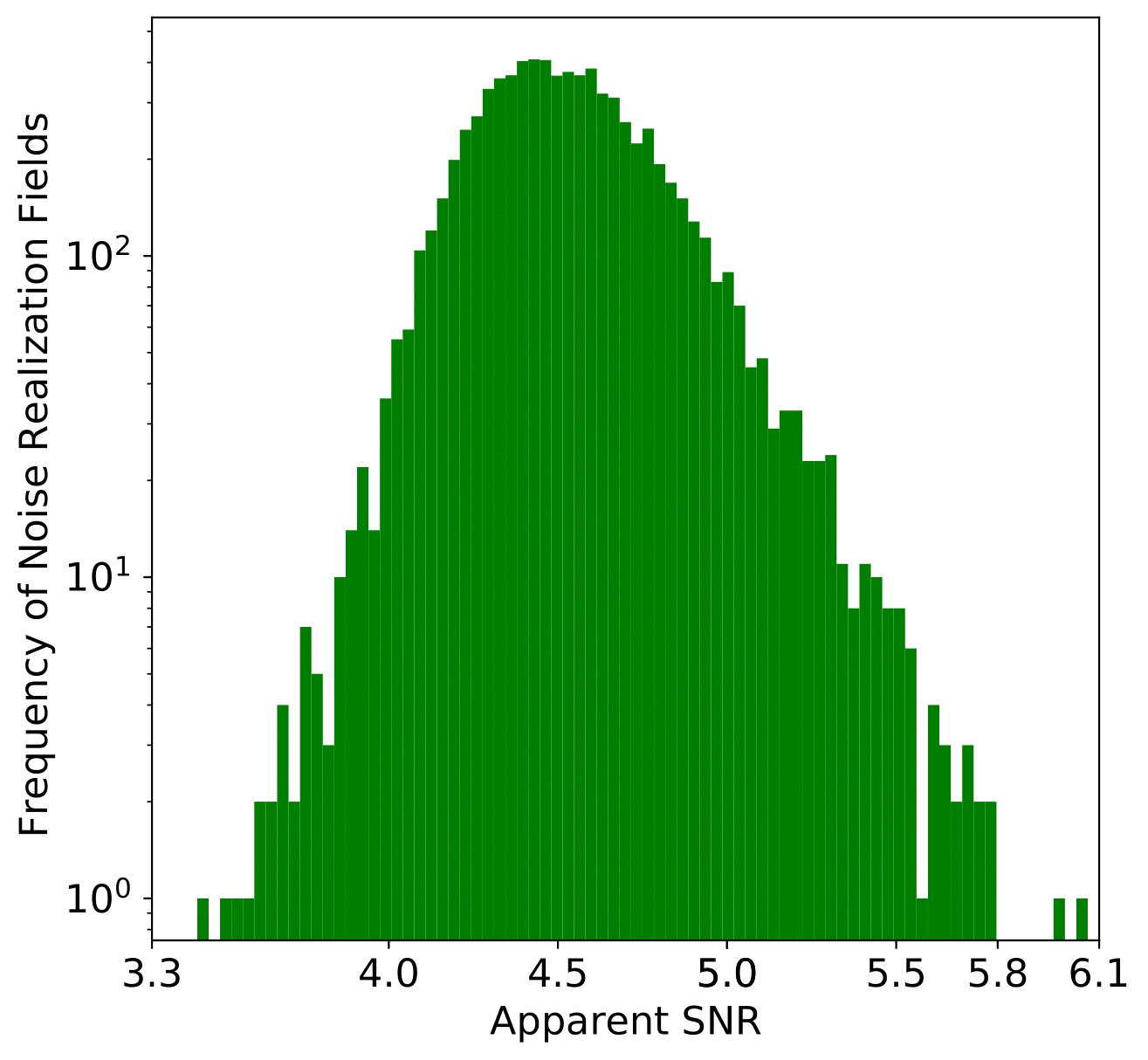}
 \caption
 {The distribution of peak surface brightness-to-noise ratio for 7800 noise realizations. The distribution is consistent with the expectations, given the total number of pixels used per image, and in total for the entire test dataset.}
 \label{fig:hist_noiserealixations}
\end{figure}

Second, we generate a dataset of 7800 noise realizations that do not contain any injected source. This dataset is used to evaluate the TN rate. To determine an apparent signal-to-noise ratio, we divide the peak surface brightness by the rms noise in each realization. The total number of noise realizations has been chosen so that we have a sufficient number of samples to test the TN rate for noise peaks at $>4 \sigma$. To ensure this, we decided to make images that are $1024\times1024$~pixels in size, from which we cropped sub-images of $256\times256$~pixels with the surface brightness peak in the centre. This results in each image having 65\,653~pixels, which for a Gaussian noise distribution should, on average, have a peak surface brightness at a significance of $4.3\sigma$, and at least one pixel detected at the $6\sigma$ level when all 7800 images are considered. In Fig.~\ref{fig:hist_noiserealixations}, we show the distribution of apparent signal-to-noise ratio of the peak surface brightness in the noise realizations. This peaks at an apparent signal-to-noise ratio of 4.4 and has at least one $6\sigma$ noise peak. Overall, the range of signal-to-noise ratios in our noise realizations is between 3.4 to 6.1. We note that the distribution is not Gaussian, with a skew towards higher signal-to-noise ratios. This is not unexpected given that the noise is correlated in the image plane for interferometric data.

\begin{table}
    \centering
\begin{tabular}{l|l|c|c|}
\multicolumn{2}{c}{}&\multicolumn{2}{c}{True Data}\\
\cline{3-4}
\multicolumn{2}{c|}{}&Source&No Source\\
\cline{2-4}
{Test Results}& Detection & TP & FP\\
\cline{2-4}
& No Detection &FN & TN \\
\cline{2-4}
\multicolumn{1}{c}{} & \multicolumn{1}{c}{Total} & \multicolumn{1}{c}{Point Sources} & \multicolumn{    1}{c}{Noise Realizations} \\
\end{tabular}
    \caption{The sample confusion matrix, showing how the TP, FP, FN and TN are defined.}
    \label{tab:sample_confusion}
\end{table}

\subsection{The performance of BLOBCAT}

As {\sc blobcat} is a source extraction algorithm that is designed to detect and catalog sources from pre-processed radio images, we first had to deconvolve the simulated visibility datasets using the {\sc tclean} task within {\sc casa} (down to a threshold of $3\sigma$). We note that this necessary step will change the simulated images being used for our comparison (dirty versus clean images), but as the underlying input model is the same for both cases, this will still allow for a proper comparison between DECORAS and a standard source detection algorithm.

{\sc blobcat} works by looking for islands of pixels that might represent a source. The signal-to-noise ratio of a given pixel is the key parameter used to detect potential sources. For this, the algorithm determines the field surface brightness and the background rms noise as the input parameters. It also requires the user to set a detection signal-to-noise ratio threshold ($T_d$) and a cut-off signal-to-noise ratio threshold ($T_f$) for flooding the islands. The detection process starts with locating all pixels that have a higher signal-to-noise ratio than $T_d$. To each pixel above this limit, an island of adjacent pixels are added that are above $T_f$. For our simulations, we consider the pair values of ($5\sigma$,$4\sigma$), ($5\sigma$,$3\sigma$), ($4.5\sigma$,$3.5\sigma$) and ($4\sigma$,$3\sigma$) for ($T_d$, $T_f$). Any detected source is then parameterized by fitting a 2-dimensional elliptical Gaussian and compared with the input model. 

\begin{figure*}
    \centering
    \includegraphics[width=\textwidth]{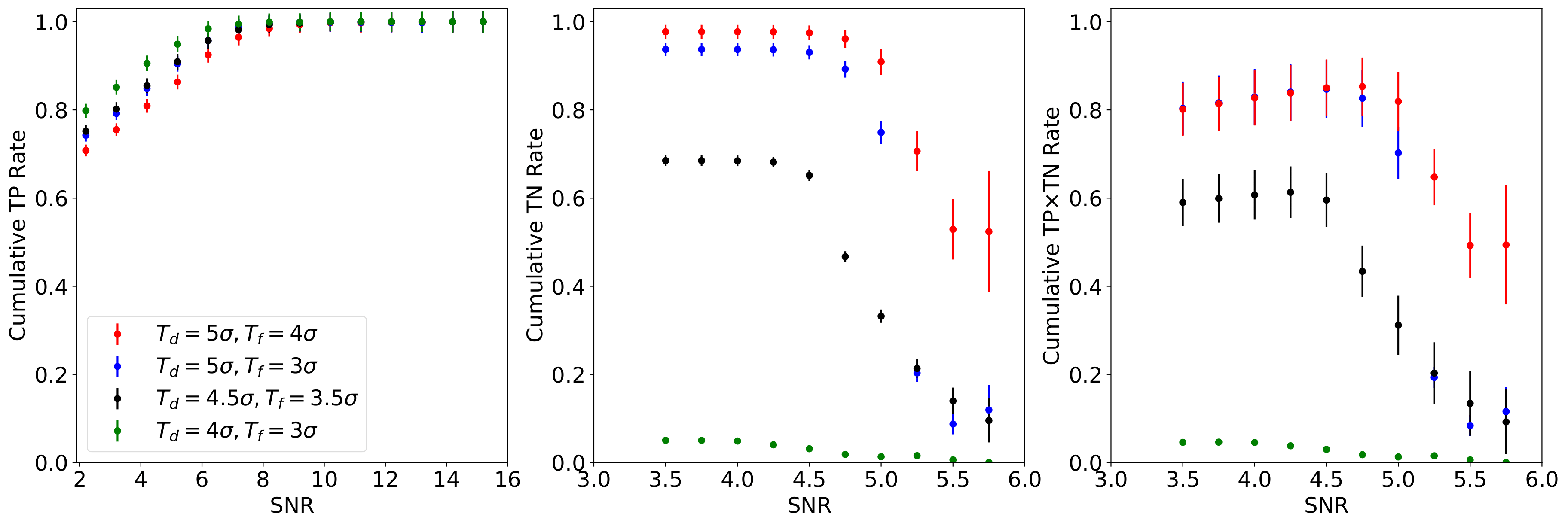}
    \caption{The cumulative true positive (TP; left), true negative (TN; middle) and ${\rm TP}~\times~{\rm TN}$ (right) rates, for {\sc blobcat} using different values for the detection ($T_d$) and flooding ($T_f$) thresholds. Setting $T_d=4\sigma$ and $T_f=3\sigma$ generates the highest TP rate (high catalog completness), but at the same time, has the lowest TN rate (low catalog purity).}
    \label{fig:blobcat_TP_TN}
\end{figure*}

In Fig.~\ref{fig:blobcat_TP_TN}, we present the cumulative TP and TN rates for {\sc blobcat} as a function of (apparent) signal-to-noise ratio. As expected, the performance of {\sc blobcat} depends on the choice of ($T_d$, $T_f$). We find that values of ($5 \sigma, 4 \sigma$) have the highest TN rate (0.98), whereas using ($4\sigma, 3 \sigma$) returns the lowest TN rate (0.05), when the full sample of noise realizations are considered. When ($T_d$, $T_f$) are set to ($4.5 \sigma,3.5 \sigma$), we find that there is a transition at a signal-to-noise ratio of 4.5, due to the fraction of FP detections decreasing and the fraction of TN detections increasing, as expected. Note that the TN rate for sources with an apparent signal-to-noise ratio below $T_d$ is high as the algorithm does not consider any blob below this significance as a potential source. We also see that the TN rate of {\sc blobcat} drops drastically for fields that have a higher apparent signal-to-noise ratio, due to the choice of thresholds that have been used.

We find that for a VLBI-like array, such as the VLBA, the point source catalogue has a TP rate of 0.91 at the $4.2\sigma$-level using a setting of ($4\sigma, 3\sigma$) for ($T_d$, $T_f$). This increases to 0.95 at the $5.2\sigma$-level and is complete at signal-to-noise ratios $>6.2$ (we define the 100 per cent completeness as the signal-to-noise ratio where the first false negative is returned). However, below this, the fraction of FN detections increases, and the overall TP rate decreases. Also, as expected, the TP rate decreases faster when a combination of higher thresholds are used. For example, the ($5\sigma, 4\sigma$) setting for ($T_d$, $T_f$) is complete at signal-to-noise ratios $>8.4$. Finally, we note that the performance of {\sc blobcat} is slightly better than the other source detection algorithms described above, with a completeness that is higher than 80~per cent at the $4\sigma$ level.

As already discussed, the settings for both $T_d$ and $T_f$ will affect the TP and TN rates determined by {\sc blobcat}. For example, using pair values of ($5\sigma, 4 \sigma$) generates a TN rate of 0.52 at the $5.5 \sigma$ level, while the ($5\sigma, 3\sigma$) setting has a TN rate of only 0.09 at the same apparent signal-to-noise ratio. Ultimately, the user's choice of $T_d$ and $T_f$ will depend on the scientific goal of the observation. For example, one might need to reach 100 per cent completeness at a specific signal-to-noise ratio, while for some other science cases a higher purity is important. In this regard, it is the combination of TP and TN rates that matters when evaluating the general robustness of a source detection algorithm. Therefore, we have calculated the combined ${\rm TP}~\times~{\rm TN}$ rate, which is also shown in Fig.~\ref{fig:blobcat_TP_TN}. We find that the TN rate is the dominant factor when comparing the performance of {\sc blobcat} with different ($T_d, T_f$) values. Also, from Fig.~\ref{fig:blobcat_TP_TN}, we see that the highest catalog completeness and purity is obtained for the ($5\sigma, 4 \sigma$) setting of ($T_d, T_f$). Therefore, we will use this setting for comparing our results with DECORAS.

\subsection{Comparing the performance of BCE and MSLE}

\begin{figure*}
 \includegraphics[width=\textwidth, trim=0cm 0cm 0cm 0cm]
 {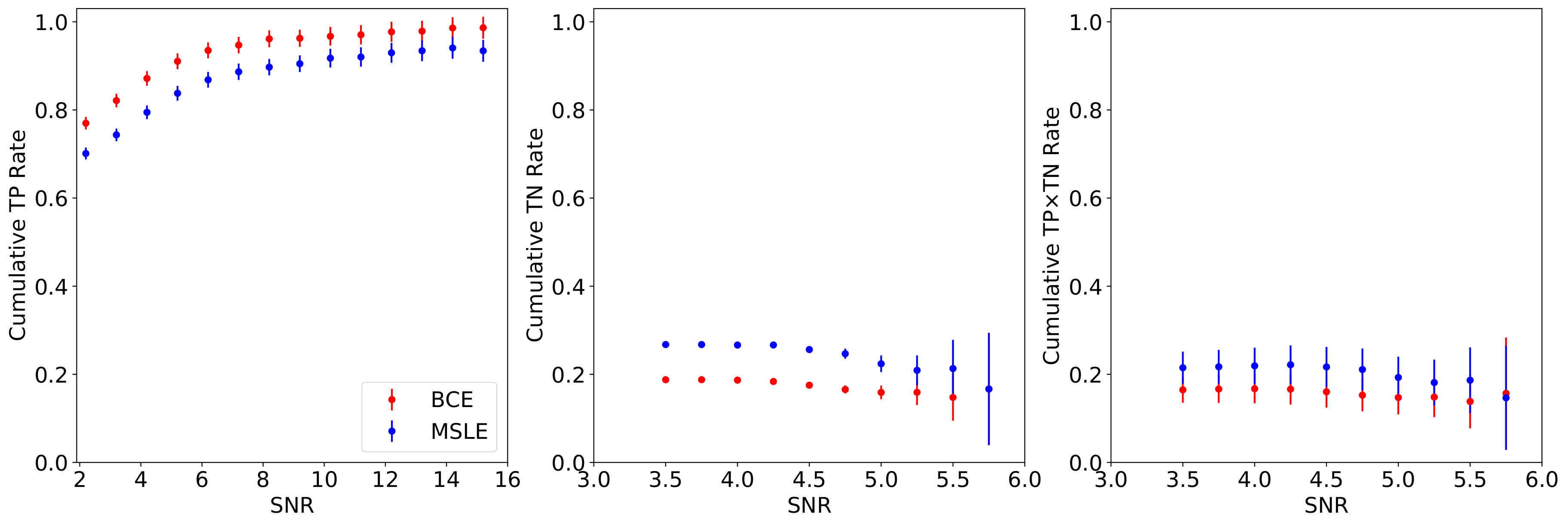}
 \caption{The cumulative true positive (TP; left), true negative (TN; middle) and ${\rm TP}~\times~{\rm TN}$ (right) rates, for the BCE (red) and MSLE (blue) loss functions when used individually. The BCE yields a higher completeness than the MSLE for all considered signal-to-noise ratios. The overall TN rates are quite low (0.19 to 0.27) when the two loss functions are used individually. Similar to the performance of {\sc blobcat}, the TN rate dominates the combined ${\rm TP}~\times~{\rm TN}$ rate.}
 \label{fig:TN_rate_bce_msle}
\end{figure*}

Fig.~\ref{fig:TN_rate_bce_msle} presents the TP, TN and ${\rm TP}~\times~{\rm TN}$ rates as a function of (apparent) signal-to-noise ratio when either the BCE or MSLE are used as the loss function in our encoder-decoder network. For the TP rate, we find that the BCE detects more genuine sources when compared to the MSLE for all signal-to-noise ratios. We find that the MSLE has a higher TN rate when compared to the BCE, which has a higher fraction of FP detections at all signal-to-noise ratios tested here. Similar to above, the ${\rm TP}~\times~{\rm TN}$ rate is dominated by the TN rate. The confusion matrices from applying the trained model using the BCE and MSLE loss functions to the test dataset are shown in Tables~\ref{tab:bce_confusion} and \ref{tab:msle_confusion}, respectively. Based on this comparison of the BCE and MSLE loss functions, we can conclude that neither provide a satisfactory performance individually.

\begin{table}
    \centering
\begin{tabular}{l|l|c|c|}
\multicolumn{2}{c}{}&\multicolumn{2}{c}{True Data}\\
\cline{3-4}
\multicolumn{2}{c|}{}&Source&No Source\\
\cline{2-4}
{Test Results}& Detection & 84.2\% & 81.2\%\\
\cline{2-4}
& No Detection & 15.8\% & 18.8\% \\
\cline{2-4}
\multicolumn{1}{c}{} & \multicolumn{1}{c}{Total} & \multicolumn{1}{c}{$8000$} & \multicolumn{    1}{c}{$7800$} \\
\end{tabular}
    \caption{The confusion matrix when only using the BCE as the loss function.}
    \label{tab:bce_confusion}
\end{table}

\begin{table}
    \centering
\begin{tabular}{l|l|c|c|}
\multicolumn{2}{c}{}&\multicolumn{2}{c}{True Data}\\
\cline{3-4}
\multicolumn{2}{c|}{}&Source&No Source\\
\cline{2-4}
{Test Results}& Detection & $78.3\% $ & $73.3\%$\\
\cline{2-4}
& No Detection & $21.7\%$ & $26.7\%$ \\
\cline{2-4}
\multicolumn{1}{c}{} & \multicolumn{1}{c}{Total} & \multicolumn{1}{c}{$8000$} & \multicolumn{    1}{c}{$7800$} \\
\end{tabular}
    \caption{The confusion matrix when only using the MSLE as the loss function.}
    \label{tab:msle_confusion}
\end{table}

\subsection{DECORAS source detection strategy}

Motivated by the results from using the BCE and MSLE loss functions individually, we now present a new strategy that is based on using both loss functions together. As described above, when the algorithm is not confident on where the source is located, more than one blob emerges in the predicted model image. In the previous section, we have considered all such low-confidence samples as non-detections when using the BCE or MSLE loss functions individually. However, using the BCE and MSLE together provides the possibility of finding the correct source, even for those fields with low signal-to-noise ratio detections. We consider a blob in a low confidence image as a detected source if both trained models, using the BCE and MSLE loss functions, agree on the existence and position of the source. A distance-threshold criteria is applied to the positions of the two detected blobs from using the BCE and MSLE individually. This threshold is defined as the maximum acceptable distance between the detected positions obtained with both loss functions, where the distance $R$ is calculated using,
\begin{equation}
\centering
    R = \sqrt{(x_{\rm MSLE} - x_{\rm BCE})^2 + (y_{\rm MSLE} - y_{\rm BCE})^2}.
\label{eq:sq_error}
\end{equation}
Here, $x$ and $y$ are the co-ordinate positions obtained with the BCE and MSLE loss functions.

\begin{figure*}
 \includegraphics[width=0.8\textwidth]{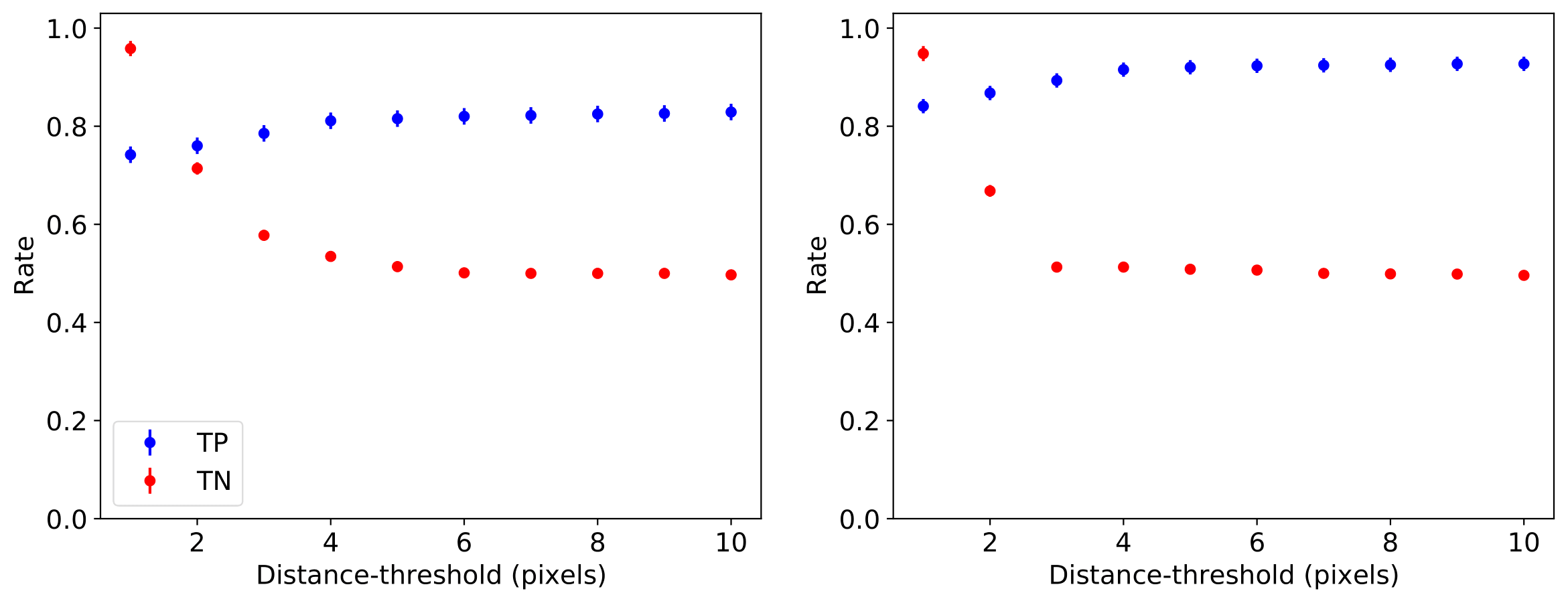}
 \caption{The true positive (TP) and true negative (TN) rates as a function of the accepted distance-threshold between the detected position using the BCE and MSLE for a signal-to-noise ratio of $>2$ (left) and $>4$ (right). Although the TP rate varies only slightly as a function of distance-threshold (0.74 to 0.82 and 0.84 to 0.93 between 1 and 10 pixels), the TN rate is highly dependent on the distance threshold.}
 \label{fig:TP_TN_squrederror}
\end{figure*}

In Fig.~\ref{fig:TP_TN_squrederror}, we show the TP and TN rates as a function of the distance-threshold $R$ when the BCE and MSLE loss functions are used together. For this, we have determined the rates for detections at a signal-to-noise ratio $>2$ and $>4$ separately. In both cases, we find that the TP rate is rather flat and only marginally changes as the distance-threshold is increased (up to 10 pixels, or 2.5 beam sizes). However, the TN rate changes drastically from 0.96 at a distance-threshold of 1 pixel to around 0.5 at 10 pixels, with the largest change occurring between 1 and 3 pixels.

\begin{figure*}
 \includegraphics[width=\textwidth] {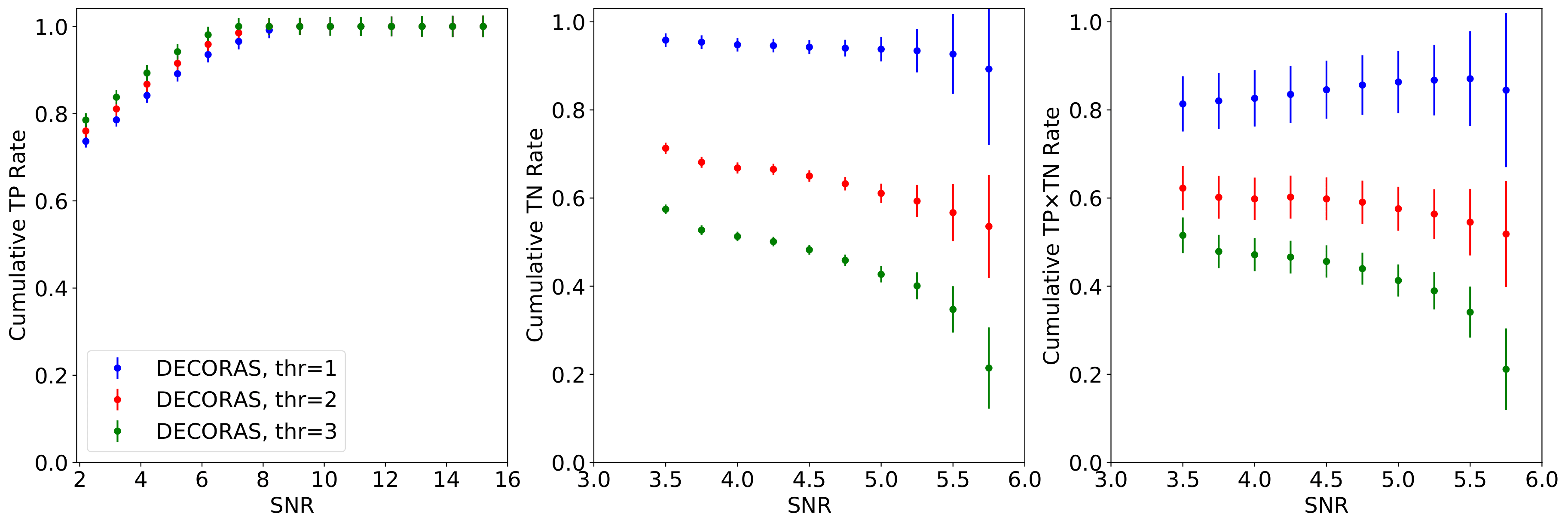}
 
 \caption{The cumulative true positive (TP; left), true negative (TN; middle) and ${\rm TP}~\times~{\rm TN}$ (right) rates, when the BCE and MSLE loss functions are used together, for distance-thresholds of 1 (blue), 2 (red) and 3 (green). The TP rates are very similar for each distance-threshold, with a value of 3 having the overall best performance. Conversely, the TN rate performance changes significantly, with a threshold of 1 having the best performance.}
 \label{fig:TP_FN_new_dst}
\end{figure*}

In Fig.~\ref{fig:TP_FN_new_dst}, we again show the TP, TN and ${\rm TP}~\times~{\rm TN}$ rates as a function of (apparent) signal-to-noise ratio, but for the case when both loss functions are used together. We also restrict our results to distance-thresholds of 1, 2 and 3 pixels, as for larger values of $R$ the TP and TN rates are essentially constant. Here, when multiple blobs have been detected by the loss functions, that is, the low confidence cases, we use a  distance-threshold to 1, 2 and 3, and for those cases with a single detected blob, we keep the distance threshold fixed to 3. From comparing with Fig.~\ref{fig:TN_rate_bce_msle}, we see that using the two loss functions together significantly improves the performance, particularly for the TN rate. We find that the TP rate has a similar behaviour for each threshold used, with slightly higher rates obtained for higher values of the distance-threshold. DECORAS is complete at signal-to-noise ratios of $>7.5$, $>6.9$ and $>6.0$ for distance thresholds of 1, 2 and 3 pixels, respectively. For the TN rates, the values are significantly higher when a distance-threshold of 1 is used, ranging from 0.97 to 0.93 between apparent signal-to-noise ratios of 3.5 and 5.5. Finally, we also see from Fig.~\ref{fig:TP_FN_new_dst} that the ${\rm TP}~\times~{\rm TN}$ rates are highest when the threshold is conservatively set to 1 for the less confidence cases. Therefore, for our comparison with {\sc blobcat}, we will consider only the results from setting the distance-threshold to 1 pixel for those samples with multiple detected blobs, and 3 for those samples with a single blob detected by both loss functions.

\begin{figure*}
 \includegraphics[width=\textwidth] {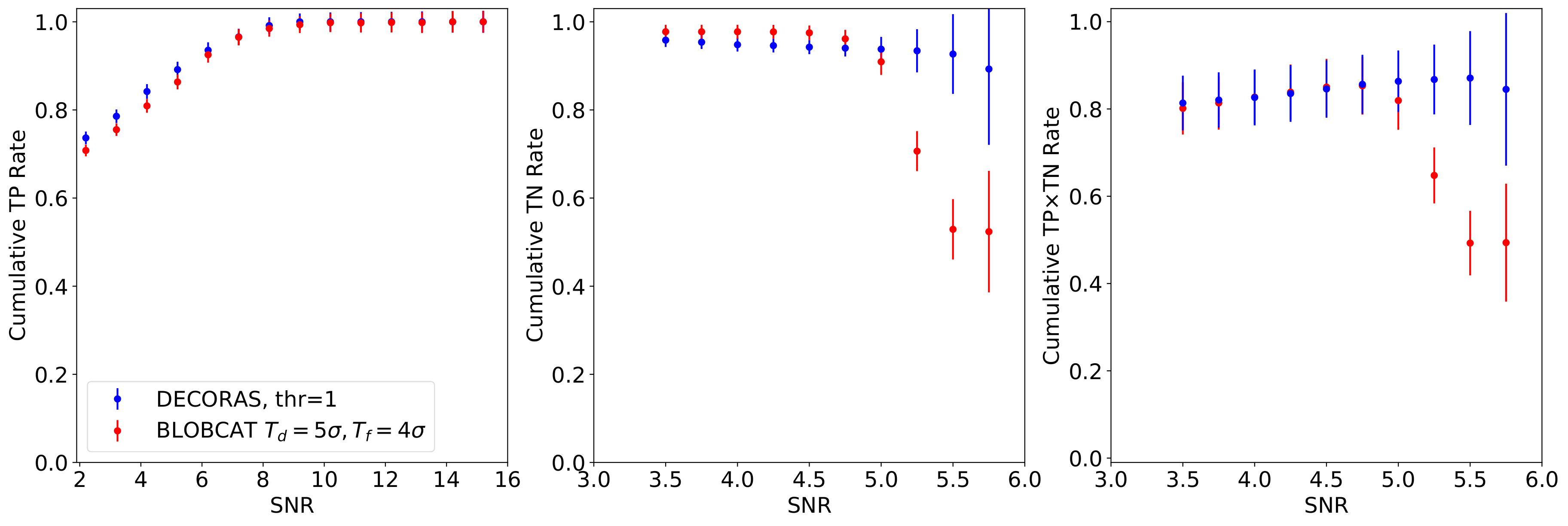}
 
 \caption{The cumulative true positive (TP; left), true negative (TN; middle) and ${\rm TP}~\times~{\rm TN}$ (right) rates for DECORAS (blue) and {\sc blobcat} (red). The TP rates are very similar for both methods, but the TN rate of DECORAS has a much better performance than {\sc blobcat} at  $>5\sigma$. This results in DECORAS having an almost factor of two better performance in combined catalog completeness and purity at signal-to-noise ratios $>5.5$.}
 \label{fig:TP_FN_best}

\end{figure*}

\subsection{Comparing DECORAS with a traditional source detection algorithm }

We present the comparison between DECORAS and {\sc blobcat} in Fig.~\ref{fig:TP_FN_best}. Overall, the TP rate (or completeness) of DECORAS is either equal to or marginally better than that of {\sc blobcat} at all signal-to-noise ratios. As described above, DECORAS is complete at the $>7.5\sigma$-level, whereas {\sc blobcat} is complete at $>8.4\sigma$. For the TN rates, {\sc blobcat} performs better at apparent signal-to-noise ratios $<5$, owing to the cut-off detection threshold, but above this, DECORAS is better, with an almost constant TN rate as there are less FP detections returned at higher signal-to-noise ratio. Therefore, we conclude that both methods have a similar completeness level, but DECORAS is expected to have a better catalog purity. This is further demonstrated in the ${\rm TP}~\times~{\rm TN}$ rate, where DECORAS out-performs {\sc blobcat} by an almost a factor of two at signal-to-noise  ratios $>5.5$.

\section{Source characterization} \label{result-characterization}

In this section, we present an analysis of how well DECORAS can recover the input source surface brightness distribution, which we parameterize as the position, size and peak surface brightness. We compare these properties with those of the input source models used to generate the test dataset discussed in Section~\ref{result-detection}. This is done by fitting two-dimensional elliptical Gaussian components to the predicted and input (true) model images of the sources.

\subsection{Recovering the source position}

\begin{figure}
\begin{center}

\begin{minipage}[b]{0.45\textwidth}
\centering
\includegraphics[width=0.8\textwidth, trim= 3cm 0cm 0cm 0cm]{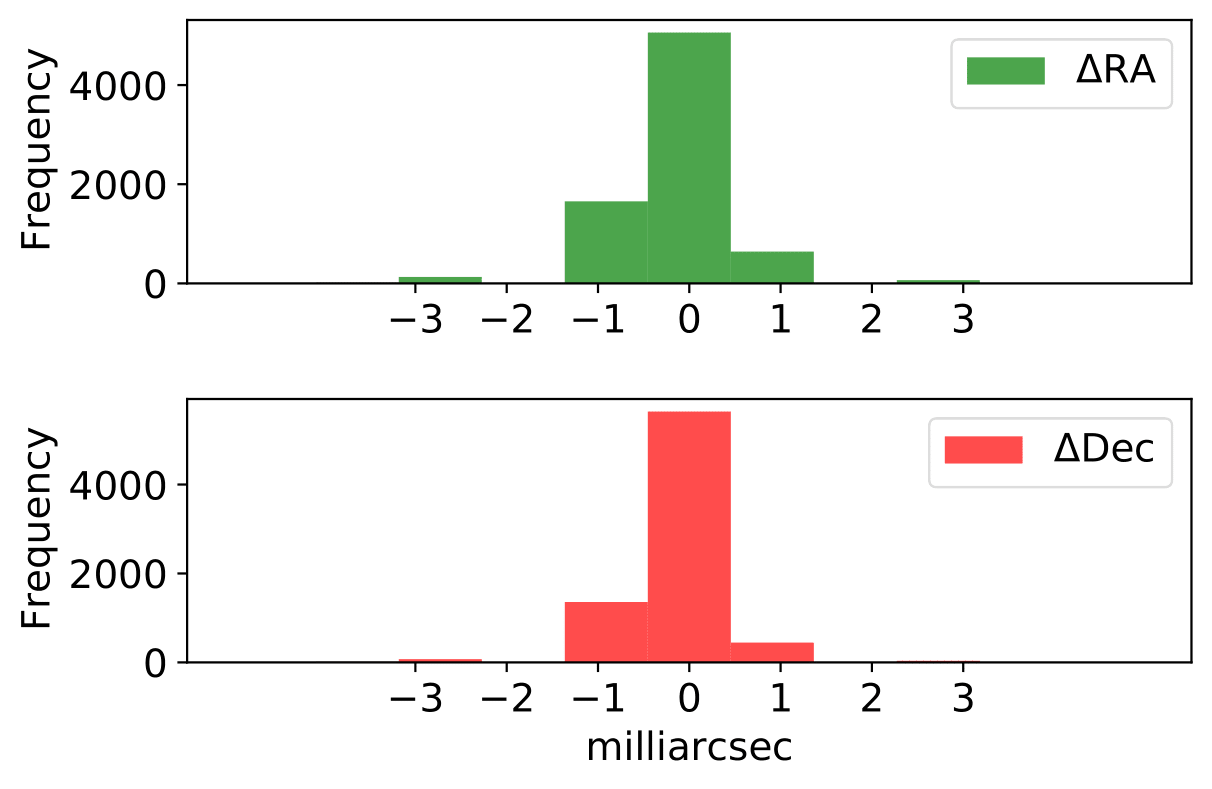}
\end{minipage}
\hfill
\begin{minipage}[b]{0.45\textwidth}
\centering
\includegraphics[width=\textwidth,trim=3cm 5cm 3cm 1cm]{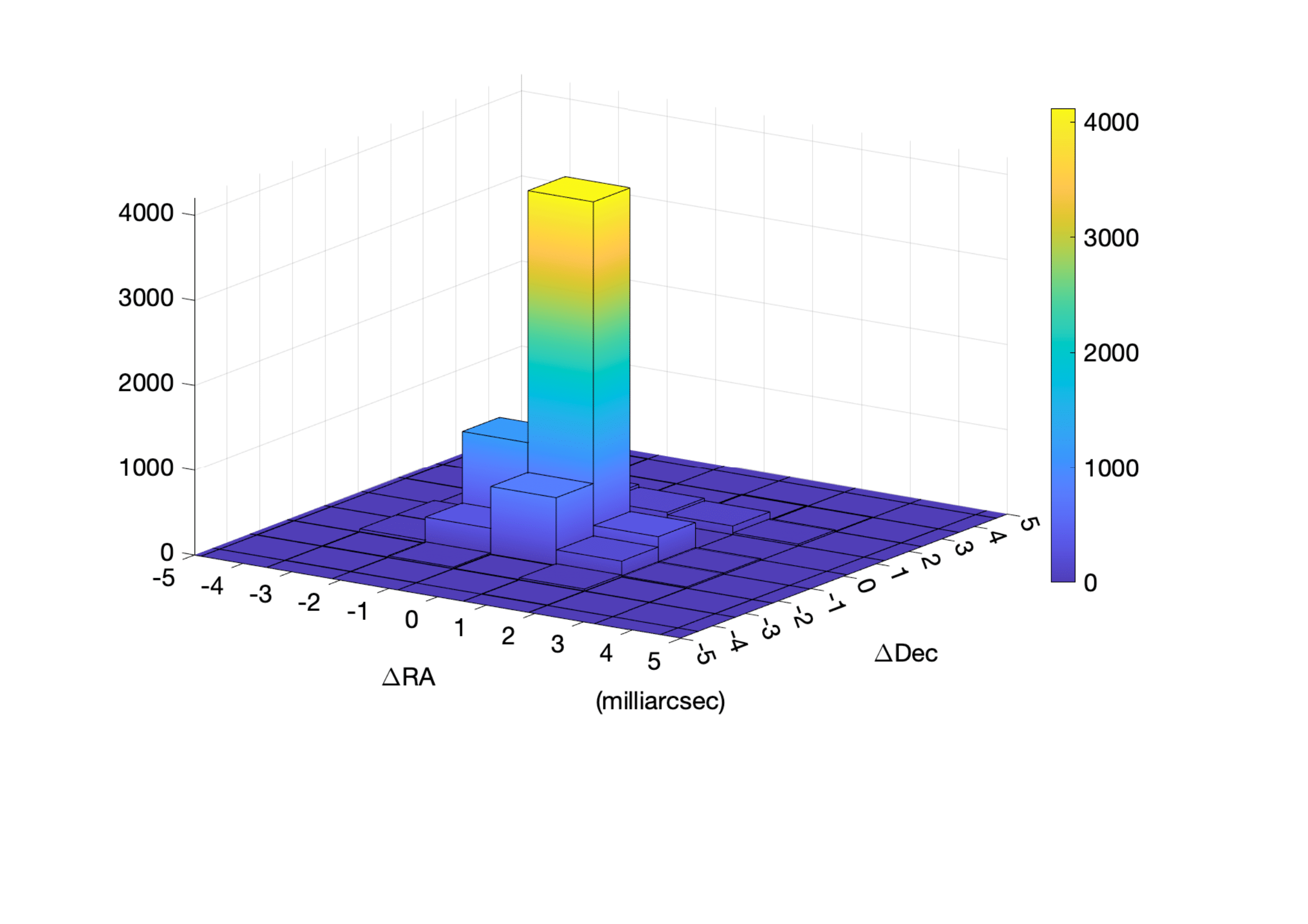}
\end{minipage}
\caption{Histograms of the relative offset, in Right Ascension  ($\Delta$RA) and Declination ($\Delta$Dec), between the predicted and input (true) position of the sources detected by DECORAS. In both directions, the mean source position is consistent with being coincident with the input model position. Note that, among all the detected sources, 0.44 and 0.40 per cent of the sources have a $\Delta$RA and $\Delta$Dec that is higher than 5 milliarcsec, respectively.}
\label{fig:delta_xy}
\end{center}
\end{figure}

Determining a reliable source position, for example, to compare the detected emission with other multi-wavelength datasets, is clearly an important aspect of any source characterization platform. In Fig.~\ref{fig:delta_xy}, we show the difference in the measured and expected position of the detected sources by DECORAS in both Right Ascension (RA) and Declination (Dec), such that, 
\begin{equation}
\Delta {\rm RA}= {\rm RA_{DECORAS}} - \rm{RA_{true}},
\end{equation}
and
\begin{equation}
\Delta {\rm Dec}= {\rm Dec_{DECORAS}} - {\rm Dec_{true}}.
\end{equation}
We find that the mean offset and standard deviation is $-0.19$ and 0.80~mas in RA, and $-0.16$ and 0.68~mas in Dec, respectively. The mean absolute offset is 0.61~mas, with a standard deviation of 0.69~mas. These results demonstrate that, the recovered source positions are consistent with the input (true) model position.

\subsection{Recovering the source structure}

We now analyze the performance of DECORAS with respect to recovering the true underlying structure of the source, which we parameterize as the source size. We have applied two different measures for this evaluation. First, we calculate the MSLE between the predicted and input (true) model image. 

\begin{figure}
 \includegraphics[width=\columnwidth]{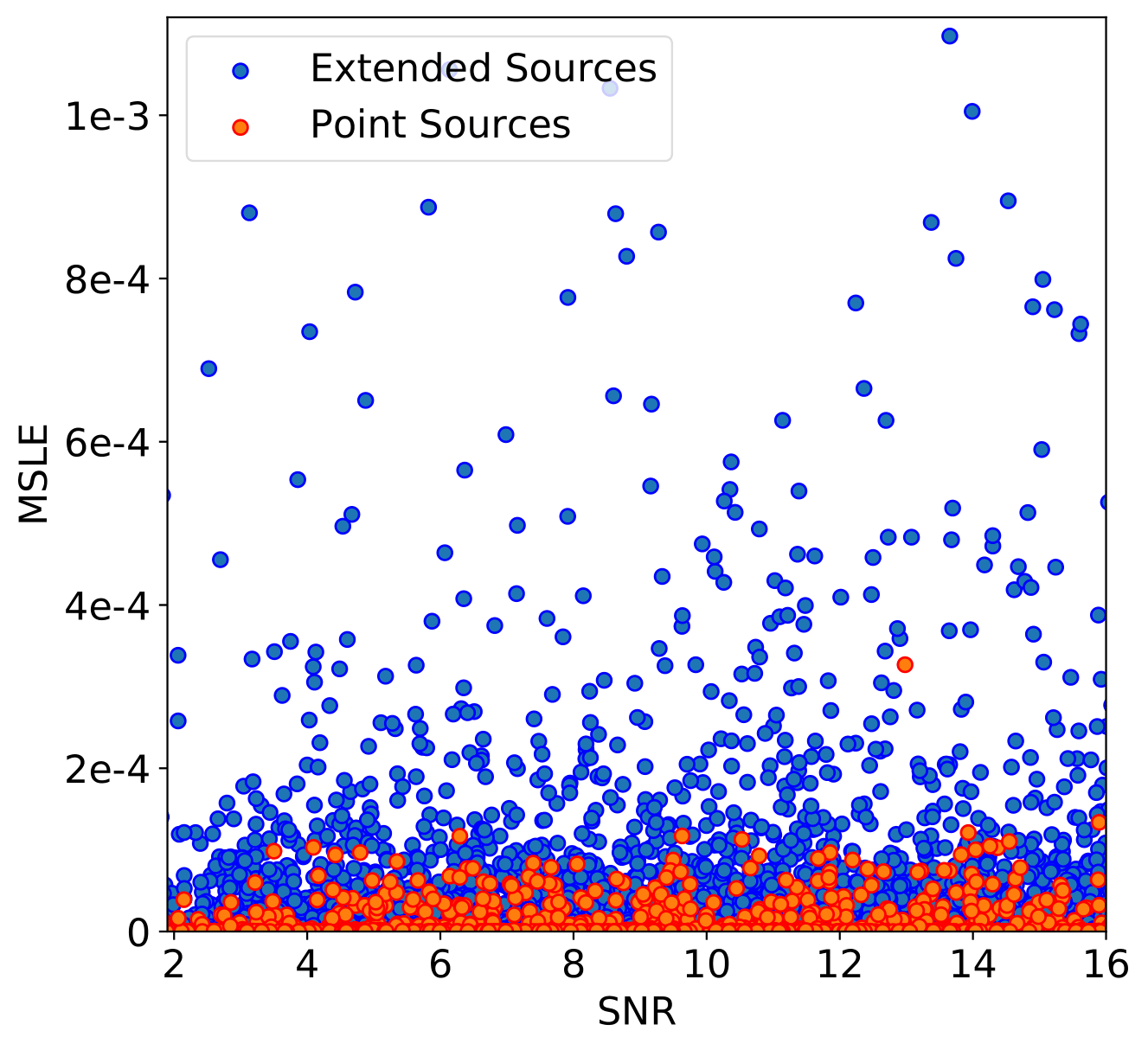}
 \caption{The MSLE between the predicted and input model images, as a function of signal-to-noise ratio for point (red) and extended (blue) sources. Overall, the MSLE values are extremely small and support our view that DECORAS reliably recovers the source structure.}
 \label{fig:msle_src_char}

\end{figure}

Fig.~\ref{fig:msle_src_char} presents the results of measuring the MSLE for the test dataset as a function of signal-to-noise ratio, for all sources detected with DECORAS. 

\begin{figure}
\begin{center}
\hspace{0.3cm} Dirty Image\hspace{1.2cm} Model Image\hspace{1cm} Predicted Image

\begin{minipage}[b]{\columnwidth}
\centering
\includegraphics[height=2.5cm]{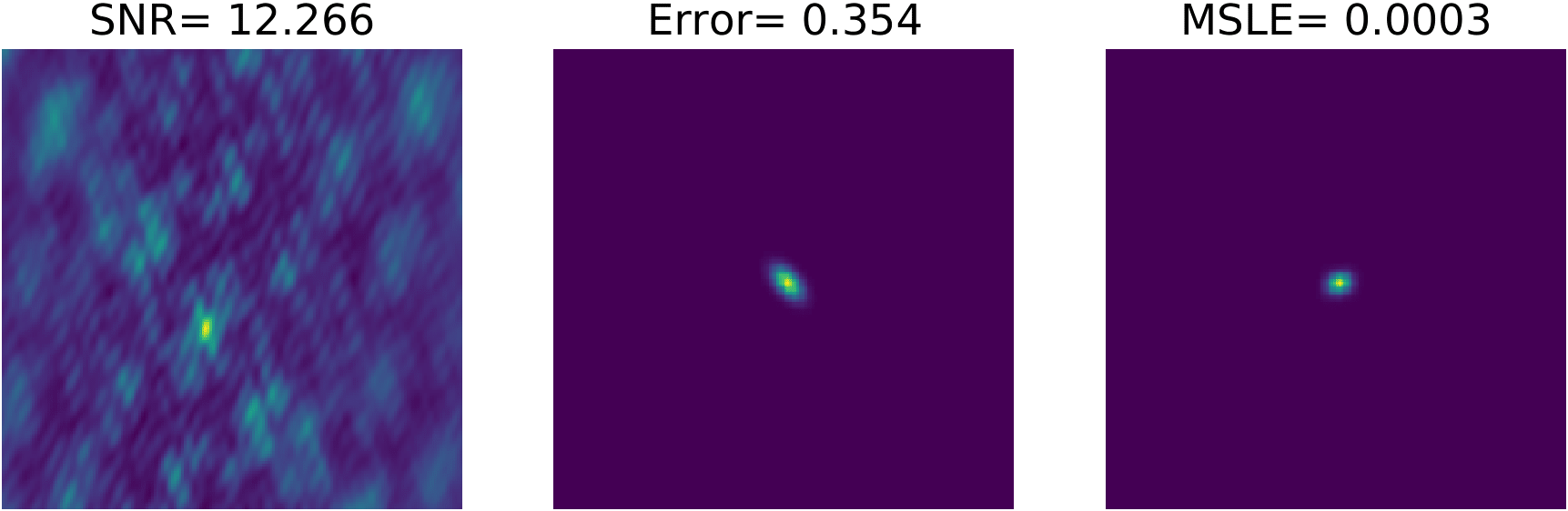}
\end{minipage}
\\
\begin{minipage}[b]{\columnwidth}
\centering
\includegraphics[height=2.5cm]{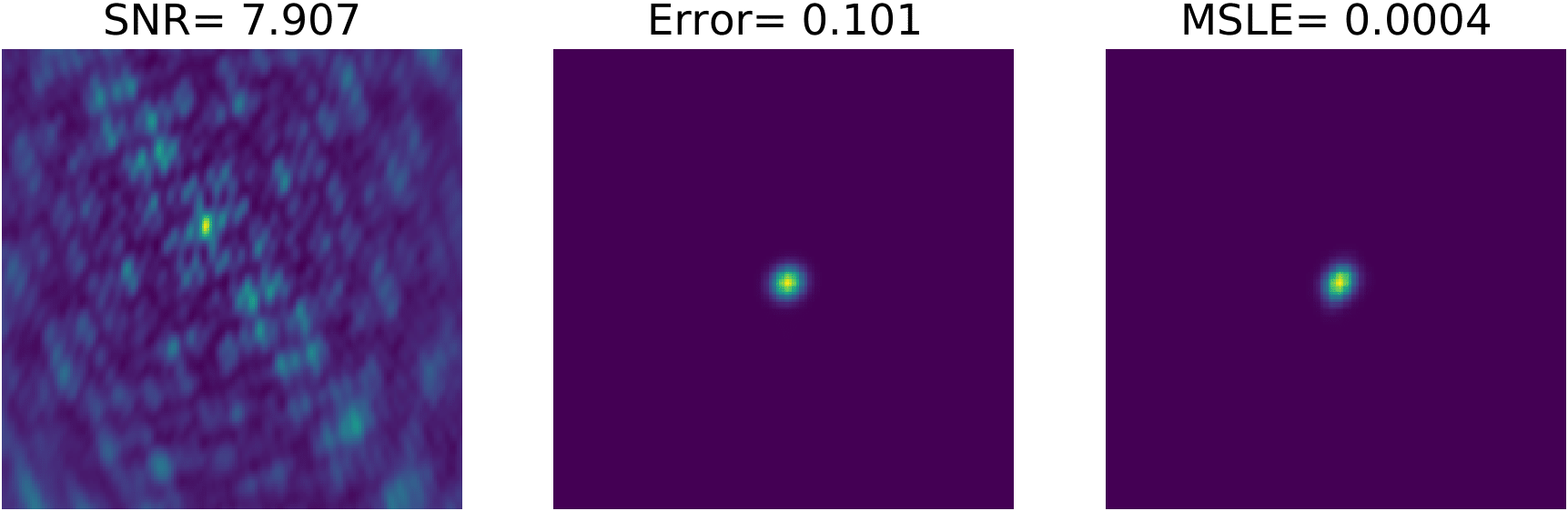}
\end{minipage}

\begin{minipage}[b]{\columnwidth}
\centering
\includegraphics[height=2.5cm]{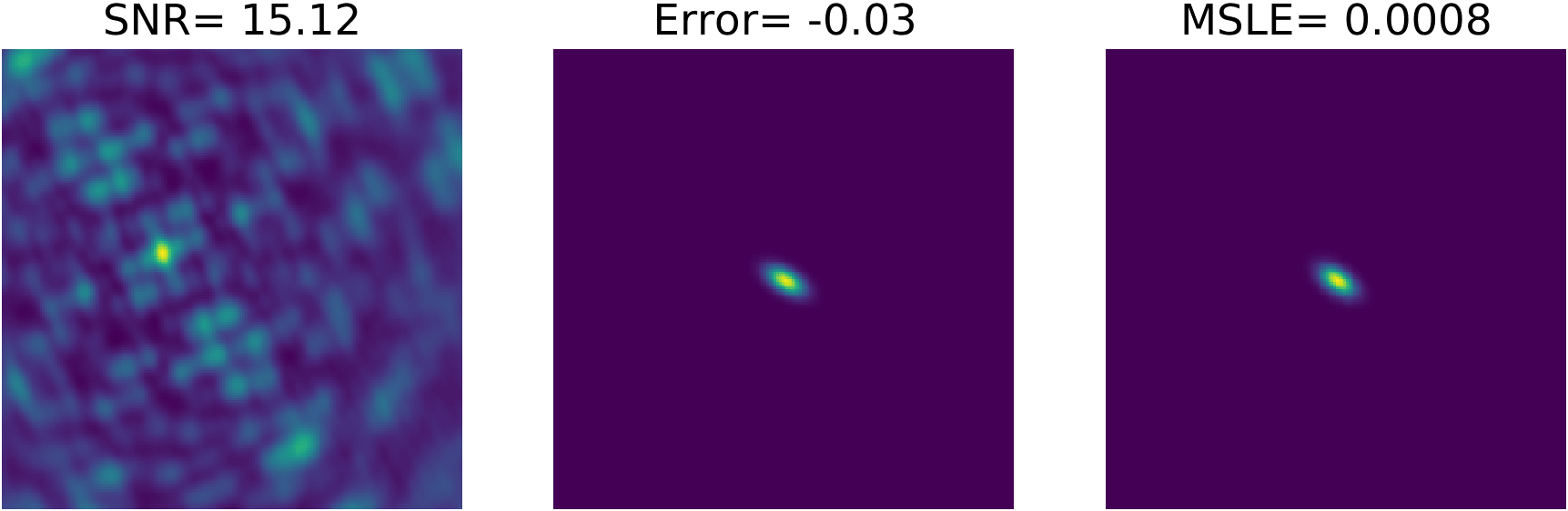}
\end{minipage} 

\begin{minipage}[b]{\columnwidth}
\centering
\includegraphics[height=2.5cm]{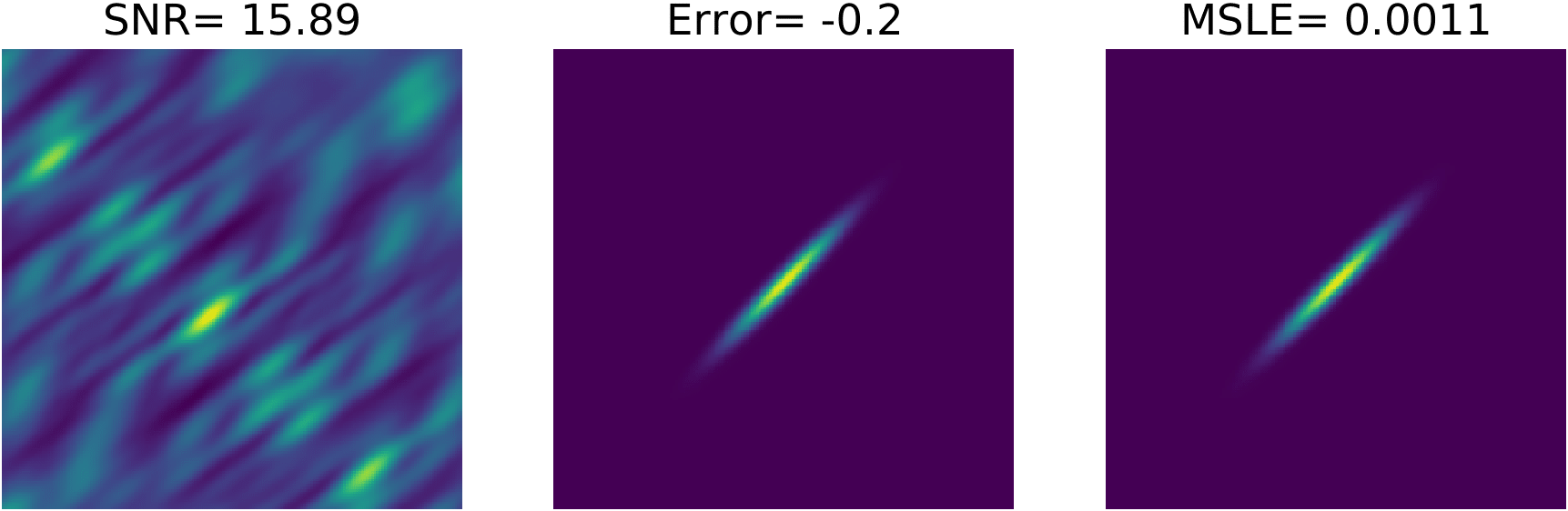}
\end{minipage}

\newpage
\caption{Examples of four sources with a high MSLE between the input (middle) and predicted (right) model image, with the associated dirty image (left) for reference. Such high values of the MSLE could be interpreted as a possible error in recovering the source structure. However, we see that the predicted and input images are comparable for these extreme cases. Each image contains $256\times256$~pixels and is equivalent to a sky-area of $320\times320$~mas$^2$.}
\label{fig:mse_higherror}
\end{center}
\end{figure}

We find that the MSLE is of order $10^{-5}$ in the majority of cases, for both point and extended sources, which translates to an acceptable source structure recovery. For example, in Fig.~\ref{fig:mse_higherror} we show four of the worst-case sources, where the MSLE is highest ($>3\times10^{-4}$). A simple visual comparison demonstrates that the model and predicted source structures are in good agreement, even for these outliers.

\begin{figure*}
 \includegraphics[width=\textwidth]{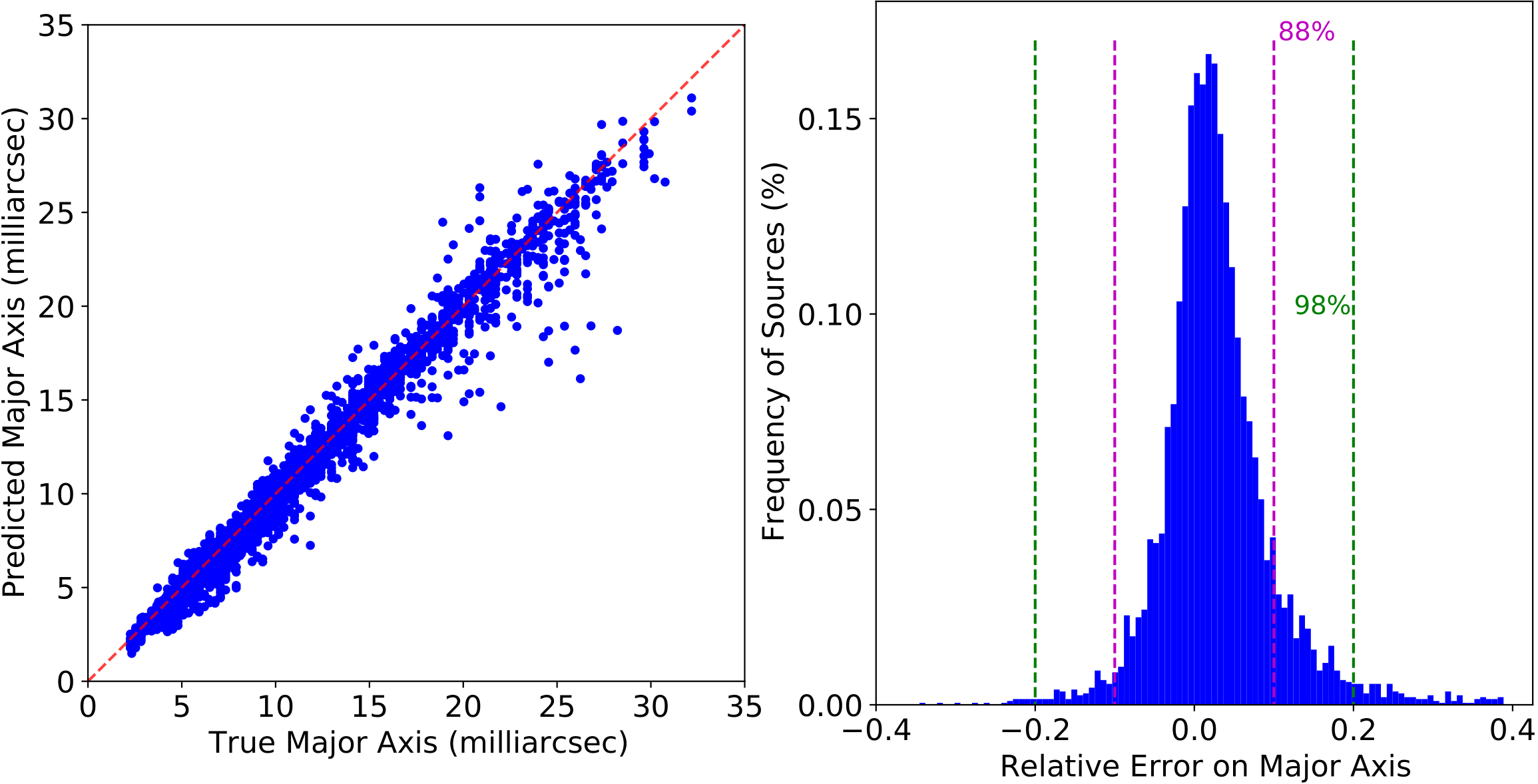}
 \caption{In the left panel, a comparison of the predicted and true major axis of each source for the entire test dataset (blue points), is presented. The red dashed line shows when they exactly agree. In the right panel, the histogram of fractional error of the predicted major axis, with respect to the input (true) model major axis, is presented. The magenta and green dashed lines show the percentage of sources within the 10 and 20 per cent fractional error bounds.}
 \label{fig:dmaj_pre_real}
\end{figure*}

Second, to quantify the performance of DECORAS in recovering the source structure, we compare the major and minor axis of the 2-dimensional elliptical Gaussian fitted to the surface brightness distribution of each predicted and input (true) model image. Fig.~\ref{fig:dmaj_pre_real} presents a comparison of the predicted and the ground truth major axis for each source, where we again see that there is good agreement. We have also calculated the relative error between the true and predicted major axis, such that,
\begin{equation}
\centering
    {\rm Relative~Error} = \frac{\rm True_{maj}-Predicted_{maj}}{\rm True_{maj}},
\label{eq:error_maj}
\end{equation}
which is also shown in Fig.~\ref{fig:dmaj_pre_real}. Again, we see that there is good agreement between the model and predicted source sizes, with the majority of the sources having almost no fractional difference. However, we see that there is still a scatter that extends up to 40~per cent in fractional error. We find that 88 (98) per cent of the sources have a fractional error of 10 (20)~per cent in major-axis size. Only 2 per cent of the sources have fractional errors in the recovered major axis of between 20 to 40~per cent. Finally, we note that there is an excess of sources towards a positive relative error, that is, the predicted major axis is smaller than the input (true) model major axis. The reason for this is not clear, but may be related to the range of beam sizes that were used during the training.

\subsection{Recovering the source peak surface brightness}
\begin{figure*}
 \includegraphics[width=0.95\textwidth]{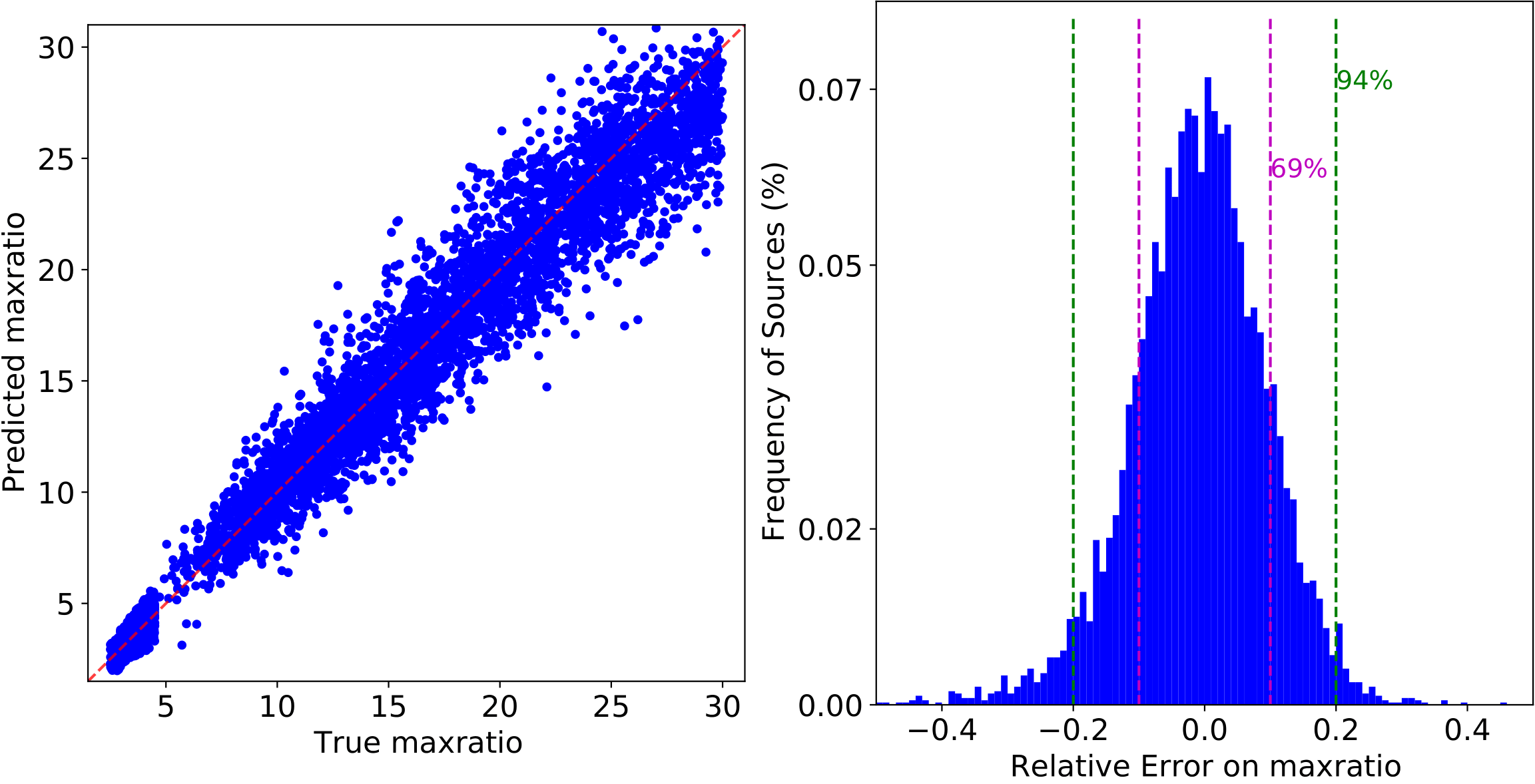}
 \caption{In the left panel, a comparison of the predicted and true maxratio of each source for the entire test dataset (blue points), is presented. The red dashed line shows when they exactly agree. In the right panel, the fractional error of the predicted maxratio, with respect to the input (true) model maxratio, is presented. The magenta and green dashed lines show the percentage of sources within the 10 and 20 per cent fractional error bounds.}
 \label{fig:predvsreal_maxratio}
\end{figure*}
Determining the absolute surface brightness is needed to measure the flux density and luminosity of the radio sources detected with DECORAS. This is important for understanding the various emission mechanisms that are at play, and for comparing the emission with other multi-wavelength datasets. Here, we use the peak surface brightness as a proxy for measuring the amplitude of the emission from the recovered sources. As discussed in Section~\ref{SBE}, the image normalization process results in losing the absolute surface brightness information. However, this is recovered using the source brightness estimator via the maxratio (see equation~\ref{eq:maxratio}) and the latent variables of Autoencoder2 (see Fig.~\ref{fig:TSNE_latent_MSLE_128}). To do this, we must first measure the accuracy of the source brightness estimator for the test dataset. We considered the KNN, XGboost, BaggingRegressor, and the RandomForest Regressor, which all are implemented using the {\sc scikit} package in Python \citep{scikit-learn}. Table~\ref{tab:regressonresults} compares the performance of these different regressors in terms of the Root Mean Squared Error (RMSE) and the $R^2$ statistic. We find that the XGboost has the best RMSE and $R^2$ statistic, when compared to the other regressors, which we adopt for the rest of our analysis. It has been implemented using the methodology outlined by \citet{friedman2000additive}; the number of gradient boosted trees is defined as 200, using a maximum depth of 7 for each of the base learners.

Fig.~\ref{fig:predvsreal_maxratio} compares the predicted and input (true) maxratio that is recovered for the test dataset. The result shows that using the latent variables and the maxatio can recover the model parameters. However, there is clearly a larger intrinsic scatter than in the case of the recovered major axis (see Fig.~\ref{fig:dmaj_pre_real}). To quantify how well the peak surface brightness is recovered, we again consider the relative error between the true and predicted parameters,
\begin{equation}
\centering
    \rm Relative~Error= \frac{True_{maxratio}-Predicted_{maxratio}}{True_{maxratio}}, 
\label{eq:error_maxratio}
\end{equation}
the results of which are also shown in Fig.~\ref{fig:predvsreal_maxratio}. We find that the predicted peak brightness is almost exactly recovered in the majority of cases, but there is also a scatter that extends to a fractional error of 60 per cent, and is independent of signal-to-noise ratio. We find that 69 (94) per cent of the sources have a fractional error on their peak surface brightness of 10 (20) per cent. Given that the absolute amplitude calibration of interferometric datasets is around 10~per cent, we conclude that our measurement errors with DECORAS will not dominate the uncertainties for the majority of the sources detected.

\begin{table}
  \centering
  \begin{tabular}{lcccc}
  \hline
 &KNN&XGBoost&Bagging&RandomForest\\
\hline
RMSE & 2.19& 2.16&2.44&2.23\\
$R^2$ (per cent) & 94.3 &94.5 &92.6 &93.7 \\
\hline 
\end{tabular}
\caption{Evaluation of the results of applying various regressors to the source surface brightness estimator. Ideally, the RMSE is as small as posisble and the $R^2$ statistic should be 100 per cent. We find that the performance of the different regressors is very similar, but that XGBoost is the best.}
 \label{tab:regressonresults}
\end{table}

\section{Discussion \& Conclusions} \label{discussion}

Source detection and characterization will always play an important role in making new scientific discoveries, particularly in the age of large synoptic survey telescopes and interferometric arrays that will operate across all observable wavelengths. The shift to larger and more complex datasets requires robust and efficient automated approaches to be developed, many of which will employ deep learning techniques. Here, we have investigated the source detection and characterization of unresolved and extended sources in a single pipeline using machine learning techniques. Our method is designed to detect sources reliably from sparse interferometric arrays, like the VLBA,  which can have highly correlated noise properties for images produced in the sky-plane domain. However, the pipeline presented here could also be used for other interferometric arrays, provided a suitable training dataset can be made. We have focused our attention on images that have not gone through a prior deconvolution process, but are instead produced from a Fourier transform of the visibility data. This was done to test how reliable such a methodology would be, as it can be extended to source detection in the visibilty plane or be used to determine residual calibration errors in the visibility data (see below).

By applying our methodology to a test dataset, which is representative of observations with the VLBA at a wavelength of 20 cm, we find that the derived catalog is 100 per cent complete down to a signal-to-noise ratio of 7.5. When we used a traditional source detection algorithm, which is applied to the same dataset, but also having gone through a de-convolution process, the completeness drops from 100 per cent at a higher signal-to-noise ratio of 8.4. This improvement in detectability provided by DECORAS is equivalent to a 25 per cent decrease in the integration time needed when compared to a traditional source detection algorithm that matches the completeness for a flux-limited sample. For example, an all-sky survey with the VLBA that reaches a similar depth to the mJIVE--20 survey would take about 5500 h to complete; applying DECORAS to such a survey could potentially save 1100 h in observing time, which is significant. Moreover, we find that DECORAS has a higher catalog purity, by almost a factor of two, when compared to a traditional source detection algorithm.

We also investigated the robustness of the source characterization using the test dataset. We found that the position of the detected sources were recovered to within 0.61 mas (0.49 pixels), with a standard deviation of 0.69 mas (0.55 pixels), from the input point source position. We also found that the peak surface brightness and size of the input sources were recovered to within 20 per cent for 94 and 98 per cent of the sources, respectively. Therefore, we conclude that the model images produced by DECORAS well represent the underlying source structure of the objects that are detected.

For recovering the source structure and source surface brightness, we had to develop a second encoder-decoder network due to the inefficiency of Autoencoder1 in recovering the source properties accurately. This is because  Autoencoder1 and its latent variables were optimized to recovering the source position rather than recovering the source properties. Our experiments show that the accuracy of recovering the source surface brightness is highly dependent on the latent variables; the better the network is trained using Autoencoder2, the lower the error will be generated on the maxratio estimation. The quality of the generated latent variables also depends on the efficiency of the network structure. Looking at the maxratio distribution, we found that there was a wide range of maxratios for extended sources. However, the number of samples with a higher maxratio was significantly lower. This will need to be considered in any future learning regression model so that enough data is provided on all maxratio bins. We noted that although the source structure is correlated with the maxratio, it is not the only parameter that affects it. The other influential parameters are unknown to us at this moment, but we expect that the PSF sidelobe structure to have a lateral effect. Also, we expect that an additional framework that separates point and extended sources using the latent variables, and trains the regressors using either the point or extended source samples will help in lowering the error on maxratio estimation.

Our deep learning architecture is rather simple with only nine convolutional layers. It was designed in this way so that there was a very short training and testing time. For example, the entire training phase of DECORAS on 50\,000 samples on a GPU node takes less than 2~h (equivalent to 7 samples per second). More complicated network structures might be able to improve on the TN and TP rates found here, but at the cost of a longer training time, and with the higher risk of overfitting. We note that the training time does not include the time needed to produce the training dataset. However, as part of this project, we have developed a pipeline to efficiently produce visibility datasets of realistic observations. This has formed the basis of an improved mock visbility dataset pipeline that generates training samples with more realistic source models, as opposed to the simple point and Gaussian source models tested here (de Roo et al., in prep.). Testing and refining DECORAS on this even more realistic dataset is our next goal before applying the algorithm to real observational data, for example, from the mJIVE--20 survey.

Also, through using more complicated source structures, the network can also be trained to identify the patterns associated with amplitude and phase errors within the visibility dataset. Currently, such errors would be absorbed in the derived source structure, which we plan to account for in a future implementation of DECORAS. This could be done by correcting the observed visibility's, or by simply accounting for the mis-match in the actual PSF from the expected PSF given the visibility sampling function; as our current implementation of DECORAS does not use the PSF or visibility sampling function for the analysis, testing (training) on a dataset with calibration errors should be a straightforward and potentially interesting next step.

Ultimately, we aim to expand this research by applying DECORAS to real observational data, such as from the mJIVE--20 survey. Due to the improved completeness and purity provided DECORAS when compared to {\sc blobcat}, we expect to detect and better characterize more sources, and to generate a more reliable catalog for the mJIVE--20 survey. This would provide a real-world test for using machine learning techniques to detect and characterize the millions of sources to be found from the next generation wide-field surveys with SKA-VLBI.

\section*{Acknowledgements}
We thank Adam Deller for making the uvfits files for the mJIVE--20 survey available for our simulations. This paper is based on research developed in the DSSC Doctoral Training Programme, co-funded through a Marie Skłodowska-Curie COFUND (DSSC 754315). JPM acknowledges support from the Netherlands Organization for Scientific Research (NWO) (Project No. 629.001.023) and the Chinese Academy of Sciences (CAS) (Project No. 114A11KYSB20170054). We would also like to thank the Center for Information Technology of the University of Groningen for their support and for providing access to the Peregrine high performance computing cluster.

\section*{Data Availability}

Upon reasonable request, the underlying data used for this article will be shared by the corresponding author.



\bibliographystyle{mnras}
\bibliography{main} 

\bsp	
\label{lastpage}
\end{document}